\documentclass[aps,superscriptaddress,eqsecnum,nofootinbib,preprintnumbers]{revtex4-2}
\usepackage{amsmath}
\usepackage{amssymb}
\usepackage{amsfonts,amsthm,bm}
\usepackage{color,xcolor}
\usepackage[greek,english]{babel}
\usepackage{titlesec}

\usepackage{graphicx,epsfig}
\graphicspath{{figures/}{./}}
\usepackage{float}
\usepackage{subfigure}
\usepackage{tikz}
\newcommand{\m}{\mu}
\newcommand{\n}{\nu}

\newcommand{\g}{\gamma}
\newcommand{\abs}[1]{\left| #1 \right|}

\newcommand{\be}{\begin{eqnarray}}
	\newcommand{\ee}{\end{eqnarray}}
\newcommand{\bea}{\begin{eqnarray}}
	
	\newcommand{\eea}{\end{eqnarray}}

\def\a{\alpha}
\def\b{\beta}
\def\g{\gamma}

\def\k{\kappa}
\def\m{\mu}
\def\n{\nu}

\def\th{\theta}

\def\Ref{\ref}

\newcommand{\beq}{\begin{equation}}
\newcommand{\eeq}{\end{equation}}
\newcommand{\bseq}{\begin{subequations}}
	\newcommand{\eseq}{\end{subequations}}

\begin{document}

\title{Scalarization of Chern-Simons Kerr Black Hole Solutions and Wormholes }

\author{Nikos Chatzifotis }
\email{chatzifotisn@gmail.com} \affiliation{Physics Department, School of Applied Mathematical and Physical Sciences,
National Technical University of Athens, 15780 Zografou Campus,
Athens, Greece.}

\author{Panos Dorlis}
\email{psdorlis0@gmail.com}
\affiliation{Physics Department, School of Applied Mathematical and Physical Sciences,
	National Technical University of Athens, 15780 Zografou Campus,
	Athens, Greece.}

\author{Nick E. Mavromatos}
\affiliation{Physics Department, School of Applied Mathematical and Physical Sciences,
		National Technical University of Athens, 15780 Zografou Campus,
		Athens, Greece.}
\affiliation{Physics Department, King's College London, Strand, London WC2R 2LS, UK.}

\author{Eleftherios Papantonopoulos}
\email{lpapa@central.ntua.gr} \affiliation{Physics Department, School of Applied Mathematical and Physical Sciences,
National Technical University of Athens, 15780 Zografou Campus,
Athens, Greece.}



\begin{abstract}
Chern-Simons gravity rotating Kerr-type black hole solutions are revisited from the point of view of their scalarization, namely examining the back reaction of pseudoscalar axion fields on the rotating geometry.  To lowest order in an angular momentum expansion, and a long range approximation for both the axion field configuration and its back reaction onto the metric, such solutions  had been discussed for the first time in the context of string theory.
In this work, we extend the analysis to give analytic expressions for slowly rotating black holes outside the horizon, which formally include an all order expansion in inverse powers of the radial distance from the centre of the black hole, which allows to approach arbitrarily close the horizon.
We also discuss in some detail the way Chern-Simons gravity violates the  energy conditions, which leads to secondary axionic hair of the rotating black hole solutions. This discussion is extended to  studies of wormhole solutions, which we construct in this article via the thin-shell procedure and we demonstrate that the presence of the axion field  enforces the two black hole solutions to be counter-rotating and for  slow rotation,   there is no backreaction of the Cotton tensor on the thin-shell of the wormhole.
\end{abstract}

\maketitle

\preprint{KCL-PH-TH/2022-{\bf 03}}

\flushbottom


\section{Introduction}

Recently there is a lot of activity studying modified gravity theories which are resulted from   high order derivative corrections inspired from string theory, from the presence of high curvature terms and from scalar fields coupled to gravity, known as scalar-tensor theories. An important question in all such theories concerns the behaviour of black hole solutions. This question acquires important physical significance, given the r\^ole of such black holes in producing gravitational waves (GW) during non-spherically symmetric coalescence or collapse~\cite{Misner}. The GW, in return, can be used as sensitive probes of the modified gravity scenarios which the black holes in question constitute solutions to, given that the features and profile of the GW in general depend on the underlying gravitational theory.

One of the basic feature of these modified gravity theories is the r\^ole of scalar fields in such black hole solutions. In most of these theories the scalar field backreacks on the geometry, the so-called process of scalarization, and dresses the black hole with hair. In the context of string theory~\cite{string,Campbell:1991kz,kerr1,kerr2,kerr3} slowly rotating Kerr-type black holes acquire scalar hair due to the presence of the so-called string model independent axion field~\cite{svrcek}, which in the context of the (3+1)-dimensional effective low energy string inspired gravitational theory, obtained after string compactification, is dual to the field strength of the spin-one antisymmetric tensor field of the massless gravitational multiplet.  On the other hand,
dilaton scalar field secondary hair in non rotating black holes in dilaton-Gauss-Bonnet (GB) string inspired gravity has been studied in \cite{kanti}. In both examples, Kerr and non rotating string inspired black holes, the existence of secondary  scalar field hair is allowed as a result of the violation of the weak energy conditions.

Gravity theories that contain higher powers of curvature are the Chern-Simons (CS) theories which are  consistent Lanczos-Lovelock gravity theories in higher than four dimensions. CS black holes  are special solutions of these theories  resulting in second order field equations for the metric with well defined AdS asymptotic solutions \cite{CSblackhole}. The CS black holes in d-dimensions were studied in \cite{Gonzalez:2010vv}  and calculating  analytically the quasinormal modes of the scalar perturbations the stability of these solutions was investigated (for a review on CS black holes see \cite{Zanelli:2005sa}).

The action for CS modified gravity is defined by the sum of the Einstein-Hilbert action and a
new parity violating, four-dimensional correction. Interest in the model spiked when it was found that string
theory unavoidingly requires such a correction to remain mathematically consistent. In the perturbative string sector, such a correction is necessary by the Green-Schwarz anomaly canceling mechanism upon four-dimensional compactification. In general, such a correction arises in the presence of Ramond-Ramond scalars due to duality symmetries. This parity violating
axion field can dress a slowing rotating black hole with (secondary) scalar hair.
One of the first works that evaluated the axion field hair that dresses a slow rotating black hole, as a result of the coupling of the axion field to a Lorentz CS term, was presented in \cite{kerr1,Campbell:1991kz}. This result suggested  that non-minimal gravitational couplings may produce interesting new effects in black hole backgrounds. This solution was further extended in \cite{kerr3} where it was found that the charges introduced by the axion hair are determined by the mass, angular momentum and gauge charges of the background rotating black hole. Allowing the axion field to be dynamical a solution describing a rotating black hole  in the small coupling slow rotation limit was found in \cite{Yunes:2009hc}. Static and rotating black string solutions  in dynamical CS modified gravity were studied in \cite{Cisterna:2018jsx,Corral:2021tww}.

In scalar-tensor theories \cite{Fujii}  hairy black hole solutions were generated when the scalar field coupled to gravity backreacts to the background metric. The asymptotic properties of the spacetime is vital for the existence of hairy black holes. One of the first hairy black hole was discussed in \cite{BBMB} but it was found that it was unstable \cite{bronnikov}. Introducing  a scale through the presence of a   cosmological constant, regular hairy black hole solutions were found in asymptotically AdS/dS spacetimes \cite{Martinez:1996gn,Martinez:2004nb,Zloshchastiev:2004ny,Martinez:2002ru,
Winstanley:2002jt,mavwin,Martinez:2006an,Kolyvaris:2009pc}. If matter interacts with black holes  non-trivial scalar field can be supported in their exterior region. Various hairy black holes were found in the context scalar-tensor Horndeski theories \cite{Horndeski} in which   matter is coupled to the Einstein tensor  \cite{Charmousis:2014zaa,Kolyvaris:2011fk,Rinaldi:2012vy,Kolyvaris:2013zfa,Babichev:2013cya,Cisterna:2014nua,Anabalon:2013}.

If the scalar field is directly coupled to curvature invariants hairy black holes can also be generated. In this case the scalar hair is maintained by the interaction with the spacetime curvature. One high curvature correction is the Gauss-Bonnet (GB) term  and to avoid topological invariant behaviour the scalar field has to be coupled to the GB term in four dimensions \cite{stringT}. These gravity theories are known as extended scalar-tensor-Gauss-Bonnet (ESTGB) theories and  were studied extensively in the literature \cite{Mignemi_1993}-\cite{Kleihaus_2016a} in a attempt to get hairy black hole solutions.
For certain classes of the scalar coupling function it was shown that we have spontaneous scalarization of black holes \cite{Doneva_2018a}-\cite{Doneva:2018rou}. The scalarization procedure were applied in  ESTGB gravity theories with various asymptotic spacetimes. in asymptotical flat spacetime black hole solutions and compact objects \cite{Silva:2017uqg}-\cite{Hunter:2020wkd} and also in AdS/dS spacetimes \cite{Bakopoulos:2018nui,Brihaye:2019gla,Bakopoulos:2019tvc,Bakopoulos:2020dfg,Lin:2020asf} and in a black hole background with different curvature topologies has been studied in \cite{Guo:2020zqm,Kiorpelidi:2022kuo}. The connections of asymptotically AdS  black holes  scalarization  with  holographic phase transitions in the dual boundary theory was studied in \cite{Brihaye:2019dck,Guo:2020sdu}. Recently the spontaneous scalarization in f(R) gravity theories  was discussed in \cite{Tang:2020sjs}.

There is a strong connection of the violation of the energy conditions with the formation of scalar hair outside the horizon of a black hole. In  both CS and GB extended scalar-tensor gravity theories \cite{kanti} there are negative contribution of the higher-curvature terms on the effective stress tensor of the theory which leads to the violation of energy conditions as we will discuss in the following. Such violation of the energy conditions also characterises the CS-axion string-inspired running-vacuum cosmology~\cite{sola1,sola}. This is a very interesting effect of the scalarization of black holes. It suggests an interplay between the curvature and matter, the spacetime geometry is deformed around the black hole because of the presence of matter which interacts with curvature.

In this work in the generic CS gravity we will look for  rotating black hole solutions  in the presence of axion fields,
 which backreact on the geometry and source the anomalous terms. Our aim is to extent previous results \cite{kerr1,kerr2,kerr3,yunes1,yunes2} in which the existing solutions truncated to long range outside the horizon, in the sense that there was no attempt to express formally the solution in terms of inverse powers of the radial distance from the centre of the black hole. Our solutions of the axionic field backreacting to a slow rotating background metric will  allow us to go arbitrarily close to the horizon of the hairy black. In this way we will be able to study the behaviour of axionic matter outside the horizon of a slow rotating Kerr-type metric and see how this matter distribution violates the energy conditions and investigate the possibility of the formation of a wormhole.

In particular, we shall  discuss in detail the violation of energy conditions that allow for the existence of secondary axion hair, following the discussion in \cite{kanti} and motivated by the violation of the energy conditions of our slow rotating Kerr-type black hole solution with axionic hair, we will investigate if a wormhole solution can be constructed. The new feature of our solution is the presence of the axionic matter or the associated Cotton tensor, and using the  gluing procedure that connects via a wormhole throat two oppositely slowly rotating Kerr black holes, we will study if the axionic matter backreacts on the whormhole geometry to the first order in the angular parameter.

The work is organized as follows. In Section \ref{sec:CSgrav}, we review the construction of the CS term
and set up the Lagrangian and equations of motion for the graviton and axion fields in CS gravity, which shall lead to the local rotating black-hole solutions. We also study the properties of the Cotton tensor which appears in the variation of the CS term, and which plays a central r\^ole in our approach. In Section \ref{sec:CSkerr} we discuss local solutions of the Einstein and axion equations of motion, of rotating Kerr-type black-hole  in the slow rotation approximation, that is to lowest order in an expansion in powers of the angular momentum or spin of the black hole. In Section \ref{sec:encond}  we discuss in detail the energy conditions of the slow rotating black hole solution  and demonstrate the violation of the NEC condition, which leads to the existence of axionic hair. In Section \ref{sec:worm}  we construct wormhole solutions via the thin shell junctioning procedure, and we discuss possible backreaction on the axionic field on the thin shell. Finally conclusions and outlook are discussed in Section \ref{sec:concl}. Some technical aspects of our approach, concerning the black hole solution and the wormhole construction, are discussed in the Appendices.

\section{Chern-Simons Gravitational Theory }
\label{sec:CSgrav}

In this Section we will review the Chern-Simons gravitational theory discussing mainly local solution of this  theory. Consider an axion matter field coupled to the gravitational CS topological term known as the Pontryagin density term ~\cite{jackiw,Alexander}. This term, henceforth denoted as $R_{CS}$, is constructed from the contraction of the Riemann tensor to the dual Riemann tensor and reads
\begin{equation}
	\label{RCS}
	R_{CS}= \frac{1}{2}R^{\mu}_{\,\,\,\nu\rho\sigma}\widetilde{R}^{\nu\,\,\,\,\rho\sigma}_{\,\,\,\mu},
	\end{equation}
where the symbol $\widetilde{(\dots)}$ denotes the dual of the Riemann tensor, defined as
\begin{equation}\label{dualriem}
	\widetilde{R}_{\alpha\beta\gamma\delta}=\frac{1}{2}R_{\alpha\beta}^{\,\,\,\,\,\,\,\,\rho\sigma}\varepsilon_{\rho\sigma\gamma\delta},
\end{equation}
with $\varepsilon_{\rho\sigma\kappa\lambda} = \sqrt{-g(x)} \, \hat \epsilon_{\rho\sigma\kappa\lambda} $  the covariant Levi-Civita  under the convention that the symbol $\hat \epsilon_{0123}=1$, {\it etc}.

The action that is giving the coupling of the axion field to the $R_{CS}$ term in presence of its kinetic energy  is given by
\begin{equation}
	S=\int d^4x \,\sqrt{-g}\,  \left[\frac{R}{2\kappa^2}-\frac{1}{2}(\partial_\mu b)(\partial^\mu b) - A\, b\,R_{CS}\right]=\int d^4x\,\sqrt{-g} \left[\frac{R}{2\kappa^2}-\frac{1}{2}(\partial_\mu b)(\partial^\mu b)\right]-\int d^4xA b\hat{R}_{CS}~,
	\label{eq:Action}
\end{equation}
where $\hat{R}_{CS}$ is the Chern-Simons term with the flat Levi-Civita tensor $\hat{\epsilon}$, $\kappa=M_{\rm Pl}^{-1}$ is the inverse of the reduced Planck mass, $b$ is a pseudoscalar denoting the axion matter field and $A$ is the coupling parameter of the axion field to the Chern-Simons term with dimension of length. Notice that the action \eqref{eq:Action} is shift-symmetric under the transformation  $b \to b + c$,  because  we have do not introduced any axion $b$ potential or mass term. Such actions are inspired from microscopic string-theory models. In the case the action \eqref{eq:Action} is derived from the corresponding low-energy string effective to quadratic order in derivatives~\cite{string,kerr1}, after appropriate string compactification, the axion field $b$ is the dual of the field strength of the spin one antisymmetric tensor field in the bosonic massless gravitational multiplet of the string, the so-called string-model independent axion \cite{svrcek}, and the parameter $A$  is fixed in terms of the string Regge slope
$\alpha'=M_s^{-2}$, where $M_s$ denotes the string scale, as follows~\cite{kerr3}
\begin{align}\label{Adefstring}
\displaystyle A=\sqrt{\frac{2}{3}}\frac{\alpha'}{48 \kappa}~,
\end{align}
and is of order $\mathcal{O}\left(\frac{M_p}{M_s^2}\right)$. The CS terms are accompanied by the axion field $b$, which, as we shall discuss below, is in the weak approximation for slowly-rotating black holes of large mass ${\mathcal M}$
compared to the Planck scale,
\begin{align}\label{masskappa}
{\mathcal M \, \kappa \gg 1}\,,
\end{align}
which we assume throughout. Hence, by selecting appropriately small angular momentum we guarantee the validity of our slow-rotation approximations in which we keep terms of up to linear order in the black hole angular momentum.

Variation of the action with respect to the metric  and the axion field yields the following equations of motion
\begin{align}
	\label{grav}
	&G_{\mu\nu}=\kappa^2 T^b_{\mu\nu}+4 \kappa^2 A C_{\mu\nu}~,\\
	\label{Axion}
	&\square b=A \, R_{CS}~,
\end{align}
where $T^b_{\mu\nu}$ is the  stress energy-momentum tensor associated with the kinetic term of a matter field,
\begin{equation}\label{stressb}
	T^b_{\mu\nu}=\nabla_\mu b\nabla_\nu b-\frac{1}{2}g_{\mu\nu}(\nabla b)^2~.
\end{equation}
The quantity
$C_{\mu\nu}$ is the Cotton tensor derived from the metric variation of $b R_{CS}$ (see Appendix \eqref{sec:Cotton}) and reads~\cite{jackiw}
\begin{equation}\label{cottdef}
	C_{\mu\nu}=-\frac{1}{2}\nabla^{\alpha}\left[(\nabla^{\beta} b) \widetilde{R}_{\alpha\mu\beta\nu}+(\nabla^{\beta} b) \widetilde{R}_{\alpha\nu\beta\mu}\right]~.
\end{equation}
We notice that, unlike the gauge anomaly term, the gravitational CS term is not topological, in  the sense of its variation with respect to the metric contributing to the total stress energy-momentum tensor. The Cotton tensor satisfies a covariant non-conservation law~\cite{jackiw,Alexander},
\begin{equation}\label{Cottcons}
	\nabla_\mu C^{\mu\nu}=-\frac{1}{4}(\nabla^{\nu}b) R_{CS}~.
\end{equation}
This result can be straightforwardly recovered using the equations of motion. Indeed, the equations of motion for the axion \eqref{Axion} are simply the Bianchi identity of the gravitational equations. Indeed, under $\nabla^{\mu}G_{\mu\nu}=0$,
where $G_{\mu\nu} \equiv R_{\mu\nu} - \frac{1}{2} \, g_{\mu\nu}\, R$ is the Einstein tensor,
and noting that $\nabla^{\mu}T_{\mu\nu}=\square b \nabla^{\nu}b$, we can immediately find that
\begin{align}\label{eqmotion}
	0= \nabla^{\mu} G_{\mu\nu}=\kappa^2\nabla^{\mu}T^b_{\mu\nu}+4\kappa^2A \nabla^{\mu}C_{\mu\nu}\,,
	\end{align}
	from which,
\begin{align}\label{eqmotion2}
	0=\kappa^2\square b\nabla^\nu b+4\kappa^2 A\,\nabla^{\mu}\,C_{\mu\nu} \,
	\overset{\eqref{Axion}}{\implies} A \, \, R_{CS}\nabla^{\nu}b=-4 A\,\nabla^{\mu}\,C_{\mu\nu} \implies \nabla^{\mu}\, C_{\mu\nu}=-\frac{1}{4}(\nabla^\nu b)\,R_{CS}~.
\end{align}
Note that the conservation of the matter stress tensor, $T^b_{\mu\nu}$ is violated, since $\nabla^{\mu}T^b_{\mu\nu}$ is
 non zero, being proportional to the CS anomalous term ({\it cf.} \eqref{eqmotion2})
\begin{align}\label{noncons}
\nabla^{\mu}\,T^b_{\mu\nu} = -4 \, A \, \nabla^{\mu}\,C_{\mu\nu} =  A \, \frac{1}{4}\, (\nabla^\nu b)\,R_{CS}~.
\end{align}
This implies an  exchange of energy between matter (the  axion field $b$) with the gravitational anomaly for spacetime backgrounds for which the latter is non trivial, such as gravitational wave perturbations~\cite{stephon,sola} or rotating black holes~\cite{kerr1,kerr2,kerr3}, of interest in the current study. There is no issue of the validity of general covariance though, given that \eqref{noncons} has been derived using self-consistently the covariant formalism. Moreover, in the specific context of string theory, the axion $b$ is itself part of the gravitational multiplet of the string, which the graviton belongs to, and hence an exchange of energy with the CS anomaly refers to interactions between members of that multiplet.\footnote{In this respect, we mention that in the cosmological model of \cite{sola}, the gravitational CS anomalies  have been assumed to characterise only the very early universe, where only fields from the string gravitational multiplet are assumed to exist as external classical fields. Matter, including radiation and chiral fermion fields, are assumed to be generated in that cosmology at the end of the inflationary period, which itself is due to a condensation of GW that lead in turn to the formation a CS anomaly condensate, contributing a de Sitter term in the effective action, which dominates in the early universe and leads to inflation without the need of inflaton fields, simply being due to gravitational non-linearities. The primordial CS terms (which could be due to both, primordial GW as well as rotating primordial black holes), though, are
cancelled by the gravitational anomalies generated by the chiral fermionic matter at the end of the inflationary period, and in this sense the post inflationary cosmological evolution does not exhibit the aforementioned axionic energy exchange with gravity.}

\section{The search for local solutions: slowly rotating Kerr-type black holes
 \label{sec:CSkerr}}

Since the axion matter field is a pseudoscalar, it imposes an axial symmetry on the underlying spacetime. Naturally, if we consider a static and spherically symmetric spacetime, the $R_{CS}$ term vanishes identically and the axion dynamics need to be absent. As such, in a spherically symmetric background, our theory is reduced to GR. However, the presence of the axion field $b$ imposes  axial symmetry and therefore we may search for rotating compact object solutions. To get an exact rotating black hole solution we will constrain ourselves to slowly rotating spacetimes.  We shall perform our analysis to leading order in the angular momentum parameter $a$ of the slowly-rotating Kerr-type spacetime,  and suppress any contributions of order $\mathcal{O}(a^2)$.

 The metric ansatz we are going to consider is the following
\begin{equation}\label{hth}
	ds^2=-G(r)dt^2+F(r)dr^2-2r^2 a \sin^2\theta W(r)dtd\phi+r^2d\Omega^2~.
\end{equation}
From the axion equation of motion (\Ref{Axion}) we find that the CS  term up to first order in $a$ yields
\begin{equation}
	\label{RCSb}
	A \, R_{CS}=-a A \cos\theta W' \frac{4 F^2 G^2+r G F'(r G'-2G)+F(-4G^2+r^2 (G')^2+2r G(G'-rG''))}{F^2 G^2r^2 \sqrt{F G}}~,
\end{equation}
while the left hand side of the axion equation (\Ref{Axion}) reads
\begin{equation}
	\label{Boxb}
	\square b=\frac{1}{F}\left[\partial_r^2+\left(\frac{2}{r}-\frac{F'}{F}+\frac{G'}{2G}\right)\right]b
+\frac{1}{r^2}\frac{1}{\sin\theta}\partial_\theta\left[\sin\theta\partial_\theta b\right]~,
\end{equation}
where prime denotes differentiation with respect to $r$. Using a separation of variables and noticing  that the right hand side of the axion equation (\Ref{Boxb}) is just the $L^2$ operator of the angular momentum of $b$, we deduce, on account of equations
\eqref{Axion}, \eqref{RCSb} and \eqref{Boxb}, that the axion field $b$ may be written as
\begin{equation}
	\label{axion1}
	b=a A u(r) P_1(\cos\theta)~,
\end{equation}
where $P_1$ denotes the Legendre polynomial of the first order, which straightforwardly cancels the angular dependence on the axionic equation of motion. We note, however, that the axion field is of minimum order $\mathcal{O}(a)$. Focusing on the Einstein equations (\Ref{grav}), we find that this implies that the $tt$ and $rr$ component of the gravitational equations of motion are satisfied in vacuum, since the corresponding stress energy tensor components are of order $\mathcal{O}(a^2)$, which we ignore in our approximation. This naturally means that
\begin{equation}
	\label{diag}
	G(r)=\frac{1}{F(r)}=1-\frac{2M}{r}~\, , \quad M \equiv G\, \mathcal M\,,
\end{equation}
with $G = 8\pi \, \kappa^{-2}$  the Newton's gravitational constant in (3+1)-dimensional spacetimes.

Therefore, any backreaction on our spacetime will be encoded in the off-diagonal component of the metric. To this end, we consider the redefinition of
\begin{equation}
	\label{offdiag}
	W(r)=\frac{2M}{r^3}+w(r)~,
\end{equation}
in order to extract the corrections, $w(r)$, on the recovered slowly rotating Hartle-Thorne spacetime \eqref{hth}, which are enforced by the axionic contribution. Using equations (\Ref{axion1}), (\Ref{diag}) and (\Ref{offdiag}), we consider the $t\phi$ component of the gravitational equations of motion (\Ref{grav}). In particular, we find that~\cite{jackiw}
\begin{equation*}
	G_{t\phi}=\kappa^2 T_{t\phi}+4 A \kappa^2 C_{t\phi}~,
\end{equation*}
where the $t\phi$ component of the Cotton tensor $C_{t\phi}$  has the form
\begin{equation}\label{cott}
	C_{t\phi}=-\frac{3 A M (r-2M)(ru'-u)}{r^5}a\sin^2\theta+\mathcal{O}(a^2)~.
\end{equation}
These lead to
\begin{equation*}
	\left(\frac{u}{r}\right)'=-\frac{1}{24 A^2 \kappa^2 M }(r^4 w')'~.
\end{equation*}
Integrating the above expression and minimizing the corresponding integration constant, which would otherwise lead to a divergent axion, we find that
\begin{equation}
	\label{axion2}
	u(r)=-\frac{r^5 w'}{24 A^2\kappa^2 M}~.
\end{equation}
Therefore, we have now fixed all the degrees of freedom of our theory except the corrections on the off-diagonal component $w$, which can be found via the axionic equation of motion, by virtue of equation (\Ref{axion2}). In particular, plugging our results on the axion equation (\Ref{Axion}) we find the following  differential equation
\begin{equation}
	\label{cor1}
	r^{11}(r-2M)w'''+2r^{10}(6r-11 M)w''+(28 r^{10}-50 Mr^9-576 A^2\kappa^2M^2 r^4 )w'+3456 A^2\kappa^2 M^3=0~.
\end{equation}
To solve the above equation we consider a series expansion on the correction function $w$. In particular, in order  the radial component of the axion field asymptotically to vanish, $w'(r)$ needs to be at least of order $\mathcal{O}(r^{-5})$ and consequently  $w(r)$ to be at least of order $\mathcal{O}(r^{-4})$, which in view of equation (\Ref{offdiag}) implies that asymptotically the leading order of $g_{t\phi}$ is that of the Hartle-Thorne spacetime~\cite{Misner}, i.e. asymptotically the spacetime coincides with the slowly rotated Kerr metric \eqref{hth}.   We define $w$ in a non-closed form as
\begin{equation}
	\label{sumcor}
	w(r)=\sum_{n=4}^{\infty}\frac{ d_n M^{n-2}}{r^n}~,
\end{equation}
where $M^{n-2}$ is introduced in order to keep the coefficients $d_n$ dimensionless. Hence, the problem of finding the geometric correction $w(r)$ has been reduced to the determination of the coefficients $d_n$. Making use of (\Ref{sumcor}), we find after some algebra that equation (\Ref{cor1}) reads
\begin{equation}
	\begin{aligned}
&3456A^2\k^2M^3-162M^7d_9+256M^7d_8+8M^2d_4 r^5+\sum_{n=-4}^{-1}\frac{M^{n+7} \left[ -(n+3)(n+6)(n+9)d_{n+9}+2(n+4)^2 (n+8)d_{n+8} \right]  }{r^n}\\
&+\sum_{n=1}^{\infty}\frac{M^{n+7} \left[ -(n+3)(n+6)(n+9)d_{n+9}+2(n+4)^2 (n+8)d_{n+8} \right] +576A^2\k^2M^{n+3}(n+3)d_{n+3} }{r^n}=0~.
	\end{aligned}
\end{equation}
In order for  the left hand site of this equation to be zero for any $r$ all the coefficients of the powers $r^n$ have to vanish. This leads us to the following equations
\begin{equation}
\begin{aligned}
&d_4=0~,\\
&256d_8-162d_9=-3456\frac{A^2\k^2}{M^4}\\
& -(n+3)(n+6)(n+9)d_{n+9}+2(n+4)^2 (n+8)d_{n+8}=0\;\;,\;\;\text{for}\;n=-1,-2,-3,-4\\
&-(n+3)(n+6)(n+9)d_{n+9}+2(n+4)^2 (n+8)d_{n+8}+576\frac{A^2\k^2}{M^4}(n+3)d_{n+3}=0\;\text{for}\;n\geq 1~.
\end{aligned}
\label{firstresultdn}
\end{equation}
 Performing the shift of $n\rightarrow n-9$, the last equation yields to
 \begin{equation}
d_n=\frac{2(n-5)^2(n-1)}{n(n-6)(n-3)}d_{n-1}+\frac{576A^2\k^2}{n(n-3)M^4}d_{n-6},\;\;\text{for}\;n\geq 10~.
\label{dnequation}
 \end{equation}
The above recursive equation requires $d_{4,5,6,7,8,9}$ to be known in order to calculate all the other coefficients. From (\ref{firstresultdn}) we find the following set of equations
\begin{equation}
	\begin{aligned}
	&d_4=d_5=0~,\\
	-28&d_7+48d_6=0~,\\
	-80&d_8+126d_7=0~,\\
	256&d_8-162d_9=-3456\frac{A^2\k^2}{M^4}~.
	\end{aligned}
\label{equations}
\end{equation}
Thus, we are left with four unknown coefficients, but with three equations. This means that we cannot completely determine the series coefficients from the differential equation (\Ref{cor1}).

However, we can extract a constraint from the weak field limit, i.e. the case where the metric does not backreact on the axionic field. This case is valid in the asymptotic region of our spacetime. Then, the metric solution is found to be the slowly rotated Kerr spacetime in a first order approximation scheme on the angular momentum. Plugging the slowly rotated Kerr metric to the axionic equation of motion (\ref{Axion}), we find the differential equation
\begin{equation}
 -2 u(r)+2(r-M)u^\prime(r)+(r^2-2 M r)u^{\prime\prime}(r) =\frac{144M^2}{r^5}~,
	\label{eq:firstDE}
\end{equation}
which is solved in the Appendix \ref{sec:appDEsol} to yield
\begin{equation}
	\label{eq:Asymptoticu}
	u(r)=-\frac{5}{4M r^2}-\frac{5}{2r^3}-\frac{9M}{2r^4}~.
\end{equation}

Consequently, the asymptotic behavior determines the first terms of the radial part of the axion. Of course, this result has to be compatible with (\Ref{axion2}), which under relation (\Ref{sumcor}) reads
\begin{equation}
u(r)=\frac{1}{24A^2\k^2M}\left[  4d_4M^2+\frac{5d_5M^3}{r}+\frac{6d_6M^4}{r^2}+\frac{7d_7M^5}{r^3}+\frac{8d_8M^6}{r^4}    \right]+\mathcal{O}(1/r^5)~.
\label{approxaxion}
\end{equation}
Performing a simple matching of the coefficients between equations (\ref{approxaxion}) and (\ref{eq:Asymptoticu}), we may deduce that the first coefficients of (\Ref{sumcor}) read
\begin{equation}\label{ddef}
	d_4=d_5=0\;\;,\;\;d_6=-5\g^2\;\;,\;\;d_7=-\frac{60\g^2}{7}\;\;,\;\;d_8=-\frac{27\g^2}{2}\;\;,\;\;d_9=0,\;\;\;\text{where}\;\;\;\gamma^2=\frac{A^2\k^2}{M^4}~.
\end{equation}
We note here that an important consequence of the above result is  that the relation (\Ref{sumcor}) produces coefficients that contain higher even orders of $\g$ by virtue of (\Ref{dnequation})
\begin{equation}
	\label{hocoeff}
	d_{10}=d_{11}=0\;\;,\;\;d_{12},\, d_{13},\, d_{14}\, \sim \, - \g^4\;\;,\; ....\;d_{21}\sim - \g^4 - \g^6\; ....,
\end{equation}
where $\sim$ denotes positive proportionality numerical coefficients, which we omit for brevity, as we only want to indicate the pertinent order in $\gamma$ for the various $d_n$ coefficients.  In Appendix \ref{sec:gamma} we demonstrate the convergence of the relevant series \eqref{sumcor} for every real value of the constant $\gamma \in \mathbb R$.
From relations \eqref{ddef}, \eqref{hocoeff} (and \eqref{dnequation}) we observe that  all coefficient are  non positive, an important result we shall make use in the following.

Then our scalarized slowly rotating Kerr metric reads~\cite{kerr1,kerr2,yunes1}
\begin{equation}\label{slowKerr}
	ds^2=-\left(1-\frac{2M}{r}\right)\, dt^2 + \frac{dr^2}{1-\frac{2M}{r}}+r^2\, d\, \Omega^2-2\, r^2 \, a\, \sin^2\theta \, W(r) \, dt \, d\phi~,
\end{equation}
where the off-diagonal correction term reads
\begin{equation}\label{amterms}
	W(r)=\frac{2M}{r^3}-\frac{A^2\kappa^2(189 M^2+120 M r+70 r^2)}{14r^8}+\mathcal{O}(A^{2n}),\,\,\,\,\,\,\text{with}\,\,\,  n = {\rm positive~integer} \geq2\, ,
\end{equation}
where the higher order terms $\mathcal{O}(A^{2n})$ can be found by  using (\Ref{dnequation}).
We remark at this point that such terms can become small perturbations of the explicitly written expressions if we assume, in a self-consistent manner with our slow rotation approximation, black hole solutions with sufficiently large mass $M$. Even in the case of \cite{sola1,sola} which  is characterised by large $A\, \kappa^2 \gg 1$, the terms of $\mathcal{O}(A^{2n})$
in \eqref{amterms} will be suppressed by inverse powers of $M$ ({\it cf.} \eqref{dnequation}), and hence can be tuned to be small by appropriately arranging $M$.

We stress that, although in this work we concentrated to lowest-order in an angular-momentum expansion $a$ for the black hole solution, our solution goes beyond the truncation considered in \cite{yunes1} as far as the incorporation of terms inversely proportional to powers of the radial distance from the centre is concerned. We include arbitrary order of such terms in an analytic fashion, as follows from the above considerations, and thus we can approach the horizon of the black hole from the exterior region arbitrarily close.\footnote{In the same spirit, the work of \cite{yunes2}, which considers $\mathcal O(a^2) $ corrections in the black-hole angular momentum, also uses truncated expressions, and various other approximations. We plan to extend our analysis to such cases as well, in the near future, going beyond these approximations.}

The corresponding axionic contribution can now be found by using (\Ref{axion1}) and (\Ref{axion2}) in the same order of $A$ as
\begin{equation}
	b=a A\cos\theta\left(-\frac{5}{4M r^2}-\frac{5}{2r^3}-\frac{9M}{2r^4}\right)+\mathcal{O}(A^m),\,\,\,\,\,\,\text{for}\,\,\, m = 2n + 1, \, \, n \in  \mathbb Z^+,
\end{equation}
with $\mathbb Z^+$ the set of positive integers.
This also yields the non-zero component of the Cotton tensor \eqref{cott} as
		\begin{align}\label{cottser}
			4\k^2A\, {C}_{t\phi} = a\frac{r-2M}{2}\sum_{n=4}^{\infty}\frac{n(n-3)d_n M^{n-2}}{r^{n+1}}\sin^2(\th)~,	\end{align}
which is negative for all $r > 2M$, that is outside the horizon, due to the fact that $d_n \le 0$ for all $n$ ({\it cf.} \eqref{ddef} and \eqref{hocoeff}).
Testing the result from the Komar integral \cite{Misner,Poisson:2009pwt} associated with the polar isometry denoted by the Killing vector $\xi=\partial_\phi$, we find that
\begin{equation}
    J=\int \star d\xi=a M~,
\end{equation}
which means that the axionic black-hole hair, proportional to $A \, a$,
is a secondary charge of the scalarized black hole, since it is not independent from the defining parameters of the black hole, i.e. the mass and the angular momentum as it happens in scalarized black hole models.

\section{Violation of energy conditions  for rotating black holes in Chern-Simons gravity with axion Hair
\label{sec:encond}}

In this Section we will discuss the violation of the energy conditions and their connection to the no-hair theorem for the axionic CS slowly rotating black holes we discussed in the previous Section. According to Bekenstein \cite{PhysRevD.51.R6608}  minimally as well as  non-minimally~\cite{mayo} coupled scalar fields that produce an energy-momentum tensor that obeys the WEC cannot be candidates for dressing a black hole with scalar hair. In the following, we present the violation of the NEC in our CS scalar gravity black hole solution (which implies the violation of all energy conditions, see the discussion in Appendix \ref{sec:energcond}), and then present a discussion on what one can conclude from such a violation as far as the secondary axionic hair is concerned, for the case of slowly rotating black holes \eqref{slowKerr}.

Consider the  action \eqref{eq:Action}, whose equations of motion \eqref{grav}, \eqref{Axion}, have been solved
above in the slowly rotating limit, with the solution to first order in the small angular momentum parameter $a$ given by
the slowly rotating Kerr-type black hole \eqref{slowKerr}. From the gravitational equations \eqref{grav}, the conserved energy-momentum tensor is
\begin{equation}
	T^{eff}_{\m\n}=\k^2T_{\m\n}+4\k^2A\,{C}_{\m\n}~.
\end{equation}
Up to first order in the angular momentum, $\mathcal O(a)$, the only contribution comes by the $t\phi$ component, which on account of \eqref{cottser}, reads
\begin{equation}
	T^{eff}_{t\phi}=4\k^2A\, {C}_{t\phi}= a\frac{r-2M}{2}\sum_{n=4}^{\infty}\frac{n(n-3)d_n M^{n-2}}{r^{n+1}}\sin^2(\th)~.
\end{equation}

As mentioned previously, due to the negative sign of all $d_n$, the contribution of the ${C}_{t\phi}$ component to the conserved energy momentum tensor is clearly  negative outside the horizon, $r>2M$. Moreover, for a stationary observer\footnote{We refer here to a stationary observer, while not to a static one as before, due to the fact the spacetime is stationary and not static. However, the definition of this observer is the same as before, i.e. a four velocity proportional to the timelike Killing vector induced to the spacetime.} the energy density is zero up to $\mathcal{O}(a)$. Thus, the existence of hair is expected to occur due to the  violation of the NEC (see Eq.~\eqref{NEC} of Appendix \ref{sec:energcond}).

To demonstrate the violation of the NEC, let us define the following future directed null vectors
\begin{equation}
	l_{\pm}^\m=\left(1,0,0,-\frac{g_{t\phi}}{g_{\phi\phi} } \pm\sqrt{\left(\frac{g_{t\phi}}{g_{\phi\phi}}\right)^2-\frac{g_{tt}}{g_{\phi\phi}}}  \right)~.
\end{equation}
The contraction with $T^{eff}_{\m\n}$, gives us
\begin{equation}
	T^{eff}_{\m\n}l_{\pm}^\m l_{\pm}^\n=\pm\frac{a (r-2M)^{3/2}}{2}\sin(\th) \sum_{n=4}^{\infty}\frac{n(n-3)d_nM^{n-2}}{r^{n+5/2}}+\mathcal{O}(a^2)~.
	\label{contraction}
\end{equation}
The above result implies: $T^{eff}_{\m\n}l_{+}^\m l_{+}^\n\leq 0$ and $T^{eff}_{\m\n}l_{-}^\m l_{-}^\n\geq 0$, where the equality holds, in both cases, at the horizon $r=2M$. Thus, the NEC is clearly violated in the exterior of the the black hole horizon, except for the poles $\theta=0,\pi$ where the contraction vanishes.

\begin{figure}[h!]
   	\includegraphics[width=12cm]{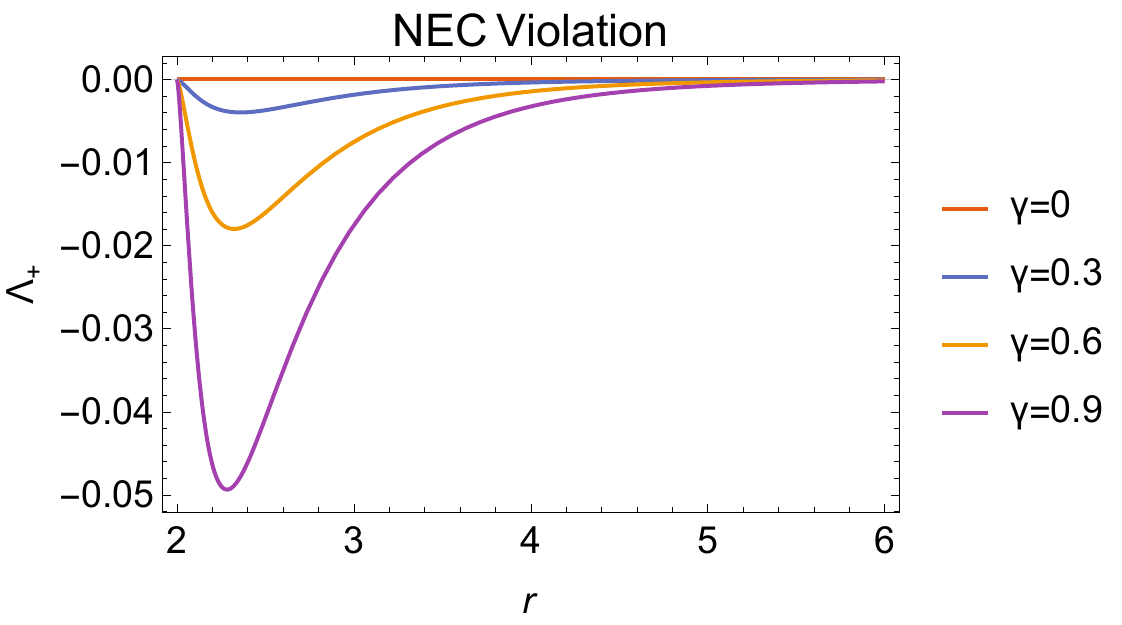}
   	\caption{Behavior of $\Lambda_+= T^{eff}_{\m\n}l_{+}^\m l_{+}^\n /a$ at the equitorial plane $\theta=\pi/2$ up to $\mathcal{O}(A^2)$ with respect to the radial coordinate.}
   	\label{fig:NECV}
   \end{figure}

This violation is attributed to the coupling of the axion with the Chern-Simons topological term, as this coupling is responsible for the appearance of the Cotton tensor in the gravitational equations of motion. Moreover, as the coupling  constant takes larger values, the NEC violation intensifies and consequently the deformation of the spacetime becomes more important. In Fig.~\ref{fig:NECV}, we illustrate the precise way by means of which this violation occurs: near the horizon, the violation takes its maximum value, while far away is negligible. So, the deformation of the Hartle-Thorne geometry is important near the horizon, while far away from it, is negligible. In addition, the NEC violation becomes stronger as we increase the value of the dimensionless parameter $\g$, which is the visualization of the statement that stronger coupling implies more deformation.

From the above considerations there seems to be a strong relation between the energy conditions and the hair of the black holes. From the gravitational equations of motion, we know that the matter fields and the underlined geometry of the spacetime are coupled, one affects the other. The energy conditions are statements that give insights on this coupling. Specifically, they shed light between properties of the matter fields and the deformation of the spacetime geometry due to the matter fields. Thus, if the existence of hair for the black holes is closely related to the violation of the energy conditions, it is closely related on how the spacetime geometry is deformed around the black holes.

Hairy black hole solutions are examples on how this spacetime deformation can be achieved. As we already discussed in the case of the CS scalar tensor gravity an axion hair has appeared, due to the coupling of the axion pseudoscalar field with the CS topological term. Moreover, this coupling is responsible for the violation of the NEC, while the violation becomes more important as the coupling becomes stronger. In addition, the violation reaches its maximum near the horizon and then it tends to zero. Thus, we can conclude that the coupling with the curvature term deforms the geometry arround the black hole in such a way that the axion hair is allowed. However, we should remark at this point that, since the axion hair appears with a secondary charge that backreacts to the metric, the independent parameters that describe the black hole are still limited to the mass, angular momentum and gauge charges, and hence the problem about the loss of information of the collapsed matter hidden inside the black hole is not solved with solely such a secondary hair.

\section{Wormhole solutions in axionic Chern-Simons theories }

\label{sec:worm}

In this Section we will discuss the possibility of generating a wormhole by the gluing   procedure that connects via a wormhole throat two oppositely slowly rotating Kerr-type  black holes we found \eqref{slowKerr} with secondary axion hair. After a  review of the thin shell formalism, we will proceed to construct the wormhole  with the aim of examining whether the presence of the CS term, responsible for the violation of the energy condition,  suffices to stabilise the wormhole.

\subsection{The thin-shell formalism}

 The gluing procedure \cite{Poisson:2009pwt,visser, Alcubierre:2017pqm,Israel:1966rt} consists of considering that our spacetime, $\mathcal{V}$, is split by an embedded hypersurface $\Sigma$ into two regions, $\mathcal{V}^+$ and $\mathcal{V^-}$. We denote the metric of $\mathcal{V}^{\pm}$ as $g_{\mu\nu}^{\pm}$ and equip it with coordinates $x^{\alpha}_\pm$. In a similar manner, we equip both sides of $\Sigma$ with coordinates $y^{i}$, where $i=(0,1,2)$. We wish to find the conditions that must be put on the metrics of each spacetime, $\mathcal{V}^{\pm}$, to ensure that the union of $g^+$ and $g^-$ forms a valid solution to the gravitational field equations. We assume that the same coordinates $y^i$ can be installed on both sides of the hypersurface and we choose $n^\mu$ to be the unit normal to $\Sigma$ pointing from $\mathcal{V}^-$ to $\mathcal{V}^+$. We will also suppose that a continuous coordinate system $x^\mu$, distinct from $x^{\mu}_{\pm}$, can be introduced on both sides of the hypersurface. These coordinates overlap with $x^{\mu}_{\pm}$ in an open region of $\mathcal{V}^\pm$ that contains $\Sigma$.

The $x^\mu$ coordinate system will be used only qualitatively on the derivation of the junctioning conditions. We note that the $\Sigma$ hypersurface is to be pierced by a congruence of geodesics that intersect it orthogonally. We take $\lambda$ to denote proper time (or proper distance) and we adjust the parametrization such that the geodesics cross the hypersurface at $\lambda=0$ and that $\lambda$ is negative in $\mathcal{V}^{-}$ and positive in $\mathcal{V}^+$. Therefore, a point $p\in \mathcal{V}^{\pm}$ identified by the coordinates $x^\mu$ is linked to $\Sigma$ by a member of the congruence, and $\lambda(x^\mu)$ is the proper time (or proper distance) from $\Sigma$ to $p$ along this geodesic. As such, a displacement away from the hypersurface along one of the geodesics may be described by
\begin{equation}
	\label{w.1}
	n^\mu=\frac{dx^\mu}{d\lambda}\implies n_{\mu}=\epsilon \partial_\mu \lambda\implies \partial_\mu=\epsilon n_\mu \partial_\lambda~,
\end{equation}
where $n_\mu$ is normalized to $n_\mu n^\mu=\epsilon=\pm 1$.
In order to perform the match of $\mathcal{V}^+$ and $\mathcal{V}^-$ through the hypersurface $\Sigma$, we need to consider the distribution function formalism. In particular, we denote $\Theta(\lambda)$ as the Heaviside distribution function for which $\partial_\lambda\Theta(\lambda)=\delta(\lambda)$, where $\delta(\lambda)$ is the Dirac distribution function. In general, we will express a quantity $A$ as
\begin{equation}
	\label{w.2}
	A(x)=A^+(x^+)\Theta(\lambda)+A^-(x^-)\Theta(-\lambda)~,
\end{equation}
where the indices $\pm$ indicate that the value of $A$ in the region $\mathcal{V}^{\pm}$. The jump of $A$ across $\Sigma$ is denoted by
\begin{equation}
	\label{w.3}
	[A]=A(\mathcal{V}^+)|_\Sigma-A(\mathcal{V}^-)|_\Sigma~.
\end{equation}
Finally, we also note the relations
\begin{equation}
	\label{w.4}
	[n^\mu]=[e^\mu_i]=0~,
\end{equation}
where $\displaystyle e^\mu_i=\frac{\partial x^\mu}{\partial y^i}$ is the pullback operator performing the mapping $\mathcal{V}^{\pm}\rightarrow \Sigma$. $[n^\mu]=0$ stems from (\Ref{w.1}) and the fact that $x^\mu$ was assumed to be continuous across $\Sigma$, while $[e^\mu_i]=0$ stems from the fact that $\Sigma$ is equipped with the same coordinates $y^i$ on both sides. Making use of the above we may extract the first junctioning condition. In particular, we note that the total spacetime metric of $\mathcal{V}$ is expressed as
\begin{equation}
	\label{w.5}
	g_{\mu\nu}=g^{+}_{\mu\nu}\Theta(\lambda)+g^-_{\mu\nu}\Theta(-\lambda)~,
\end{equation}
while its derivative reads
\begin{equation}
	\label{w.6}
	\partial_\rho g_{\mu\nu}=\partial^+_\rho g^{+}_{\mu\nu}\Theta(\lambda)+\partial^-_\rho g^{-}_{\mu\nu}\Theta(-\lambda)+\epsilon n_\rho\delta(\lambda)[g_{\mu\nu}]~.
\end{equation}
In the construction of the gravitational tensors we will encounter terms of the form $(\partial g)^2$, which will yield terms of the form $\delta^2(\lambda)$, which are ill-defined. This means that, in order for the gluing procedure to work, we need the first junctioning condition to read
\begin{equation}
	\label{w.7}
	[g_{\mu\nu}]=0~.
\end{equation}
We note that the total spacetime metric admits the decomposition $g_{\mu\nu}=h_{\mu\nu}+\epsilon n_\mu n_\nu$, where $h_{\mu\nu}$ is the first fundamental form of the hypersurface defined in the $x^\mu$ coordinate system. As such, the first junctioning condition may be reexpressed via (\Ref{w.4}) as $[g_{\mu\nu}]=[h_{\mu\nu}]$, which by virtue of the pullback operators $h_{ij}=h_{\mu\nu}e^\mu_i e^\nu_j$ reads
\begin{equation}
	\label{w.8}
	[h_{ij}]=0~.
\end{equation}
As a general guideline, we note that any quantity $A$ that is expressed in the form of (\Ref{w.2}), will yield a non-trivial $[A]$ term when differentiated, which needs to be tested for consistency in the construction of the equations of motion.
By virtue of (\Ref{w.6}) and (\Ref{w.7}), the affine connection is easily found to be
\begin{equation}
	\label{w.9}
	\Gamma^{\alpha}_{\,\,\,\beta\gamma}=\Theta(\lambda) \Gamma^{+\alpha}_{\,\,\,\beta\gamma}+\Theta(-\lambda) \Gamma^{-\alpha}_{\,\,\,\beta\gamma}~,
\end{equation}
and its derivative similarly reads
\begin{equation}
	\label{w.10}
	\partial_\delta \Gamma^{\alpha}_{\,\,\,\beta\gamma}=\Theta(\lambda)\partial_\delta \Gamma^{\alpha}_{+\beta\gamma}+\Theta(-\lambda)\partial_\delta \Gamma^{\alpha}_{-\beta\gamma}+\epsilon \delta(\lambda)[\Gamma^{\alpha}_{\,\,\,\beta\gamma}]n_{\delta}~.
\end{equation}	
Therefore, the Riemann tensor reads
\begin{equation}
	\label{w.11}
R^{\alpha}_{\,\,\,\beta\gamma\delta}=\Theta(\lambda)R^{\alpha}_{+\beta\gamma\delta}+\Theta(-\lambda)R^{\alpha}_{-\beta\gamma\delta}
+\delta(\lambda)\mathfrak{R}^{\alpha}_{\,\,\,\beta\gamma\delta}~,
\end{equation}
where
\begin{equation}
	\label{w.12}
\mathfrak{R}^{\alpha}_{\,\,\,\beta\gamma\delta}=\epsilon\left([\Gamma^{\alpha}_{\,\,\,\beta\delta}]n_{\gamma}
-[\Gamma^{\alpha}_{\,\,\,\beta\gamma}]n_{\delta}\right)~.
\end{equation}
We see that the Riemann tensor is properly defined as a distribution, but the $\delta$ function represents a curvature singularity at $\Sigma$. The second junction condition usually seeks to eliminate this term in GR, but we also have the contribution of the Cotton tensor to consider. We note that a physical interpretation can nevertheless be given to the singularity. It corresponds to the curvature induced by a thin-shell of matter, which is used to construct the wormhole by the gluing of the two spacetimes. A second note is that, since $\mathfrak{R}$ is the difference of connections, it forms a tensor. As such, it needs to be expressed in a tensorial quantity, which is calculated in Appendix \ref{sec:curvglue}, and reads
\begin{equation}
	\label{w.13}
	\mathfrak{R}^{\alpha}_{\,\,\,\beta\gamma\delta}=(n_\beta n_\gamma [K^{\alpha}_{\delta}]-n_\beta n_\delta [K^\alpha_\gamma]-[K_{\beta\delta}]n^\alpha n_\gamma+[K_{\beta\gamma}]n^\alpha n\delta)~.
\end{equation}
In a similar fashion, following the analysis in Appendix \ref{sec:curvglue}, we find that
\begin{equation}
	\label{w.14}
	G_{\mu\nu}=G_{\mu\nu}^+\Theta(\lambda)+G_{\mu\nu}^-\Theta(-\lambda)+\delta(\lambda)\mathfrak{G}_{\mu\nu}~,
\end{equation}
where
\begin{equation}
	\label{w.15}
	\mathfrak{G}_{\alpha\beta}=-\epsilon ([K_{\alpha\beta}]-[K]g_{\alpha\beta})-[K]n_\alpha n_\beta~.
\end{equation}
We may now move on to the stress-energy tensor components of our theory.

\subsection{The junctioning conditions of Chern-Simons gravity}

We will firstly consider the trivial case of $T_{\mu\nu}$. We note that the axionic field is to be expressed by virtue of (\Ref{w.2}) as
\begin{equation}
	\label{CS.1}
	b=b^+\Theta(\lambda)+b^-\Theta(-\lambda)~.
\end{equation}
Naturally, in order for $\displaystyle T_{\mu\nu}=\nabla_\mu b\nabla_\nu b-\frac{1}{2}g_{\mu\nu}(\nabla b)^2$ to not contain any terms of the form $\delta(\lambda)^2$, we need to impose that
\begin{equation}
	\label{CS.2}
	\partial_\mu b=\partial^+_\mu b^+\Theta(\lambda)+\partial^-_\mu b^-\Theta(\lambda)+\epsilon \delta(\lambda)[b]\implies	[b]=0~,
\end{equation}
which means that the jump of the axionic field vanishes identically.
Therefore, it is quite easy to deduce that
\begin{equation}
	\label{CS.3}
	T_{\mu\nu}=T^{+}_{\mu\nu}\Theta(\lambda)+T^-_{\mu\nu}\Theta(-\lambda)~,
\end{equation}
which means that the stress-energy tensor of the axion does not contribute to equations of motion of the thin-shell, i.e. the terms with $\delta(\lambda)$. It is in this sense that in order for the wormhole construction to work in GR, a thin-shell of matter must be inserted by hand in order for the second junctioning condition, $\mathfrak{G}_{\alpha\beta}=0$, to be violated.

We will now analyse the effects of  the Cotton tensor. We firstly note that, by virtue of $\displaystyle \abs{g}=-\frac{1}{4}\hat{\varepsilon}^{\mu\nu\rho\sigma}\hat{\varepsilon}^{\alpha\beta\gamma\delta}\, g_{\mu\alpha}\, g_{\nu\beta}\, g_{\rho\gamma}\, g_{\sigma\delta}$ and the metric (\Ref{w.5}), we find that
\begin{equation}
	\label{CS.4}
	\abs{g}=\abs{g^+}\Theta(\lambda)+\abs{g^-}\Theta(-\lambda)~,
\end{equation}
which yields
\begin{equation}
	\label{CS.5}
	\varepsilon_{\mu\nu\rho\sigma}=\varepsilon^+_{\mu\nu\rho\sigma}\Theta(\lambda)+\varepsilon^-_{\mu\nu\rho\sigma}\Theta(-\lambda)~.
\end{equation}
To simplify the calculations, we rewrite the Cotton tensor as
\begin{equation}
	\label{CS.6}
	C_{\mu\nu}= \, \nabla_\alpha D^{\alpha}_{\,\,\,\mu\nu}~,
\end{equation}
where
\begin{equation}
	\label{CS.7}
	D^{\alpha}_{\,\,\,\mu\nu}=\left(-\frac{\partial^\xi b}{2} (\widetilde{R}^{\alpha}_{\,\,\,\mu\xi\nu}+\widetilde{R}^{\alpha}_{\,\,\,\nu\xi\mu})\right)~,
\end{equation}
where by virtue of relations (\Ref{w.11}), (\Ref{w.12}) and (\Ref{CS.5}), the dual Riemann reads
\begin{equation}
	\label{CS.8}
	\widetilde{R}^{\alpha}_{\,\,\,\mu\xi\nu}
	=\Theta(\lambda)\widetilde{R}^{\alpha}_{+\mu\xi\nu}+\Theta(-\lambda)\widetilde{R}^{\alpha}_{-\mu\xi\nu}
+\delta(\lambda)\mathfrak{R}^{\alpha}_{\,\,\,\mu\gamma\delta}\varepsilon^{\gamma\delta}_{\,\,\,\,\,\xi\nu}~.\\
\end{equation}
Therefore, the tensorial quantity of (\Ref{CS.7}) is decomposed as
\begin{equation}
	\label{CS.9}
	D^{\alpha}_{\,\,\,\mu\nu}=	D^{\alpha}_{+\mu\nu}\Theta(\lambda)+D^{\alpha}_{-\mu\nu}\Theta(-\lambda)+\delta(\lambda)\left(-\frac{\partial^\xi b}{2}\mathfrak{R}^{\alpha}_{\,\,\,\mu\gamma\delta}\varepsilon^{\gamma\delta}_{\,\,\,\,\,\xi\nu}+(\mu\leftrightarrow\nu)\right)~.
\end{equation}
Since the Cotton tensor is found by the divergence of $D^{\alpha}_{\,\,\,\mu\nu}$, the calculation will yield components of the form $\partial_\lambda \delta(\lambda)$, which are ill-defined. In order to avoid this and make the procedure work, we impose that
\begin{equation}
	\label{CS.10}
	\left(-\frac{\partial^\xi b}{2}\mathfrak{R}^{\alpha}_{\,\,\,\mu\gamma\delta}\varepsilon^{\gamma\delta}_{\,\,\,\,\,\xi\nu}+(\mu\leftrightarrow\nu)\right)=0~.
\end{equation}
Computing now the divergence of $D^{\alpha}_{\,\,\,\mu\nu}$ to calculate the Cotton tensor, we immediately find that
\begin{equation}
	\label{CS.11}
	C_{\mu\nu}=C^+_{\mu\nu}\Theta(\lambda)+C^-_{\mu\nu}\Theta(-\lambda)+\epsilon\delta(\lambda) n_\alpha[D^{\alpha}_{\,\,\,\mu\nu}]=C^+_{\mu\nu}\Theta(\lambda)+C^-_{\mu\nu}\Theta(-\lambda)+\delta(\lambda) \mathfrak{C}_{\mu\nu}~,
\end{equation}
Let us now compute $\mathfrak{C}_{\mu\nu}$. We note that
		\begin{align*}
			[\mathcal{D}^{\alpha\mu\nu}]&=\mathcal{D}^{\alpha\mu\nu}_+|_\Sigma-\mathcal{D}^{\alpha\mu\nu}_-|_\Sigma\\
			&=\left(\frac{-\partial^+_\beta b^+}{2}\widetilde{R}_+^{\alpha\mu\beta\nu}+\frac{-\partial^+_\beta b^+}{2}\widetilde{R}_+^{\alpha\nu\beta\mu}-\frac{-\partial^-_\beta b^-}{2}\widetilde{R}_-^{\alpha\mu\beta\nu}-\frac{-\partial^-_\beta b^-}{2}\widetilde{R}_-^{\alpha\nu\beta\mu}\right)|_\Sigma\\
			&=-\frac{1}{2}\left[\partial_\beta b\, \widetilde{R}^{\alpha\mu\beta\nu}\right]-\frac{1}{2}\left[\partial_\beta b\, \widetilde{R}^{\alpha\nu\beta\mu}\right]~,
		\end{align*}
which by $[n_{\alpha}]=0$, yields that
		\begin{equation}
			\label{CS.12}
			\mathfrak{C}_{\mu\nu}=-\frac{1}{2}\left[\partial^\beta b\, \widetilde{R}^{\alpha}_{\,\,\,\mu\beta\nu}n_\alpha\right]-\frac{1}{2}\left[\partial^\beta b\, \widetilde{R}^{\alpha}_{\,\,\,\nu\beta\mu}n_\alpha\right]~.
		\end{equation}
We note that in contrast to $\mathfrak{G}_{\mu\nu}$, $\mathfrak{C}_{\mu\nu}$ is not tangent to the hypersurface. In particular,
			\begin{equation*}
				\mathfrak{C}_{\mu\nu} n^\mu=-\frac{1}{2}\left[\partial^\beta b\, \widetilde{R}^{\alpha}_{\,\,\,\nu\beta\mu}n^\mu n_\alpha\right],\qquad 	\mathfrak{C}_{\mu\nu} n^\nu=-\frac{1}{2}\left[\partial^\beta b\, \widetilde{R}^{\alpha}_{\,\,\,\mu\beta\nu}n^\nu n_\alpha\right]~.
			\end{equation*}
By making use of the symmetries of the dual Riemann, the only way that $\mathfrak{C}_{\mu\nu}$ is tangent to the hypersurface is if $\partial_\beta b=f(x)n_\beta$, i.e. the derivative of axion field is normal to the hypersurface. However, by virtue of the fact that the axionic field of our solution has angular dependence as well, this also can't be the case. Therefore, we need to insert a thin-shell with a non-trivial matter stress-tensor on our action in order for the gluing to work, since there is no way to cancel out the normal components of $\mathfrak{C}_{\mu\nu}$.\footnote{This means that one needs to consider the full Gauss-Codazzi equations, \cite{Poisson:2009pwt}, in order to completely determine the dynamics of the thin-shell.}
			
Therefore, we deduce that the modified Lanczos equations \cite{Israel:1966rt} in CS gravity read
			\begin{equation}
				\label{CS.13}
				\mathfrak{G}_{\mu\nu}-4 A\kappa^2 \mathfrak{C_{\mu\nu}}=\kappa^2 S_{\mu\nu}~,
			\end{equation}
where $S_{\mu\nu}$ is the surface stress-energy tensor of the thin-shell expressed in the total spacetime coordinates $x^\mu$.

\subsection{The gluing of the axionic black holes}

		We move on to the construction of thin-shell wormholes by slowly rotating axionic black holes. The metrics on each manifold $\mathcal{V}^{\pm}$ read according to our axionic solution (\Ref{slowKerr}),
		\begin{equation*}
			ds^2_{\pm}=-f_{\pm} dt_{\pm}^2+\frac{dr^2_{\pm}}{f_{\pm}}-2r^2_{\pm}\sin^2\theta_{\pm}a_{\pm} W_{\pm}dt_{\pm}d\phi_{\pm}+r^2_{\pm}d\Omega_{\pm}^2=-f_{\pm} dt_{\pm}^2+\frac{dr^2_{\pm}}{f_{\pm}}+r_{\pm}^2d\theta^2_{\pm}+r^2_{\pm}\sin^2\theta_{\pm}(d\phi_{\pm}-a_{\pm} W_{\pm} dt_{\pm})^2
		\end{equation*}
		where the second equality is derived from the fact that $\mathcal{O}(a^2)$ is a trivial contribution by virtue of the black hole being in the slow rotation limit. We can simplify calculations by transforming the metric in a co-rotating frame by
		\begin{equation}
			\label{g.1}
			d\psi=d\phi-a W(R) dt~,
		\end{equation}  which yields that
		\begin{equation}
			\label{g.2}
			ds^2_{\pm}=-f_{\pm} dt_{\pm}^2+\frac{dr^2_{\pm}}{f_{\pm}}+r_{\pm}^2d\theta^2_{\pm}+r^2_{\pm}\sin^2\theta_{\pm}(d\psi_{\pm}-a_{\pm}\Delta W_\pm dt_\pm)^2~,
		\end{equation}
		where $\Delta W=W(r)-W(R)$.	Therefore, the total space coordinates, $x^\mu_{\pm}$ are $(t,r,\theta,\psi)_{\pm}$. 	We consider now the embedding map $t_{\pm}=T_{\pm}(\tau), r_{\pm}=R_{\pm}(\tau)$,	where $\tau$ is the proper time of an observer comoving with the hypersurface.
		Then, each corresponding induced metric reads
		\begin{align}
			\label{g.3}
			ds^2_-|_\Sigma=-\left[F(R_-)\dot{T}_-^2-\frac{\dot{R}_-^2}{F(R_-)}\right]d\tau^2+R_-^2(\tau)d\Omega^2~,\\
			\label{g.4}		ds^2_+|_\Sigma=-\left[F(R_+)\dot{T}_+^2-\frac{\dot{R}_+^2}{F(R_+)}\right]d\tau^2+R_+^2(\tau)d\Omega^2~,
		\end{align}
			where $F_{\pm}=1-2M/R_\pm(\tau)$ and the dot denotes differentiation with respect to $\tau$. In order to perform the matching between the two induced metrics, we set that the induced metric $h_{ij}$, reads
		\begin{equation}
			\label{g.5}
				ds^2_\Sigma=-d\tau^2+R^2(\tau)d\theta^2+R^2(\tau)\sin^2\theta d\psi^2~,
		\end{equation}
		i.e., the hypersurface coordinates $y^i$, are $(\tau,\theta,\psi)$ where, naturally $ \theta=\theta_+=\pi-\theta_-$. The matching of the bulk metrics on the hypersurface implies that $R_{+}=R_{-}=R$ for the throat radius, while $\dot{T}_{\pm}$ is found to satisfy
		\begin{equation}
			\label{g.6}
				F_{\pm}\dot{T_{\pm}}=\beta_{\pm}=\sqrt{\dot{R}_{\pm}^2+F_{\pm}}~.
		\end{equation}
We consider that the two glued black holes are of the same mass, which yields that $F_+=F_-\implies T_+=T_-\implies \beta_+=\beta_-$. We note that by virtue of the co-rotating coordinate system, we can not yet have a relation for the rotation of the two black holes.
		
We have to derive  the pullback operators, $e^\mu_i$ and the unit normal one-form, $n_\mu$.
		In particular, the four-velocity of an observer comoving with the throat of the wormhole reads
		\begin{equation}
			\label{g.7}
			e^\mu_{\tau}=\left(\frac{\beta}{F},\dot{R},0,0\right)~.
		\end{equation}
		This means that, as expected, the observer comoving with the hypersurface  has vanishing $\psi$-velocity.
		On the other hand, the unit normal vector can be readily found to yield
		\begin{equation}
			\label{g.8}
			n_\mu^{\pm}=\pm\left(-\dot{R},\frac{\beta}{F},0,0\right)~.
		\end{equation}
		We note that $n_\mu e^\mu_\tau=0$ and $n_\mu n^\m=1$ are indeed verified. Under (\Ref{g.7}) and (\Ref{g.8}), we may find that the extrinsic curvature of the hypersurface $\displaystyle K^{\pm}_{ij}=-n_\alpha^\pm\left(\frac{\partial^2 x^\alpha}{\partial y^i\partial y^j}+\Gamma^{\alpha}_{\,\,\,\beta\gamma}\frac{\partial x^\beta}{\partial y^i}\frac{\partial x^\gamma}{\partial y^j}\right)$ reads
		\begin{equation}
			\label{g.9}
				K^{\pm}_{ij}=\pm
				\begin{pmatrix}
					-\frac{\dot{\beta}}{\dot{R}}&0&-\frac{a_\pm R^2\sin^2\theta W'}{2}\\
					0&R\beta&0\\
					-\frac{a_\pm R^2\sin^2\theta W'}{2}&0&R\sin^2\theta\beta
			\end{pmatrix}
		\end{equation}
where prime denotes differentiation with respect to $R$.
	
In order  to compute the contribution of the Cotton tensor on the hypersurface, we need to compute the terms of (\Ref{CS.12}), $\displaystyle -\frac{1}{2}\partial^\beta b\widetilde{R}^{\alpha}_{\,\,\,\mu\beta\nu}n_\alpha+(\mu\leftrightarrow\nu)$, on each spacetime and then calculate their difference, if any, on the hypersurface. In order to do that, we recall that the axionic solution is of the form
		\begin{equation}
			\label{g.10}
			b_\pm= a_\pm A u_{\pm}\cos\theta_{\pm}~,
		\end{equation}
where $u$ is the radial component. In order for $[b]=0$ to be  satisfied, under the constraint that $\theta_+=\pi-\theta_-$, we immediately find that
		\begin{equation}
			\label{g.11}
			a_+=-a_-~,
		\end{equation}
 which means that the two black holes need to be counter-rotating. The counter-rotating property has been previously considered in \cite{Mazharimousavi:2014tba,Krisch:2009zs}. Our reasoning here, however, is rather different. The result is enforced a priori, in order for the axion to not contain any discontinuities on the throat. We note that in a faster rotating paradigm, the axion would contain higher orders in the angular momentum and Legendre polynomials of higher rank. This could in principle differentiate the result of $[b]\implies a_+=-a_-$ to a much more complicated expression with richer physical properties.
		 Under the constraint that the two black holes are actually counter-rotating, and then we find that the Cotton tensor contribution on the hypersurface  vanishes identically, i.e.
		 \begin{equation}
		 	\label{g.12}
		 	\mathfrak{C}_{\mu\nu}=0~.
		 \end{equation}
	 	Therefore, by considering the addition of a thin-shell of matter, which enables the construction of the wormhole via the Lanczos equations, we find that, in the slowly rotating paradigm, the surface stress-energy tensor yields
	 	\begin{equation}
	 		\label{g.13}
	 			S_{ij}=-\frac{\epsilon}{\kappa^2}([K_{ij}]-[K]h_{ij})~,
	 	\end{equation}
	which is the familiar expression of the GR case.
	  Calculating the surface stress-energy tensor yields that
	  \begin{equation}
	  	\label{g.14}
	  		S_{ij}=\frac{-\epsilon}{\kappa^2}
	  		\begin{pmatrix}
	  			4\frac{\beta}{R}&0&0\\
	  			0&-R^2\left(
	  			\frac{2\beta}{R}+\frac{2\dot{\beta}}{\dot{R}}\right)&0\\
	  		0&0&-R^2\sin^2\theta\left(
	  			\frac{2\beta}{R}+\frac{2\dot{\beta}}{\dot{R}}\right)
	  	\end{pmatrix}
	  \end{equation}
  	 by virtue of the fact that two black holes are counter-rotating. Any subsequent analysis on this case will not yield any fruitful results, since (\Ref{g.14}) is the same surface stress-energy tensor as in the Schwarzschild case~\cite{Misner,thorne1,thorne2, Poisson:1995sv}. We may conclude that the  backreaction of the Cotton tensor on the thin-shell is not giving any new result in the case of the slowly rotating black hole to first order in the angular parameter $\mathcal O(a)$, and thus one needs to consider faster rotating paradigms in order to test for the stability of the resulting thin-shell wormhole.\footnote{We mention that the contributions of the variation of the CS term on the hypersurface ({\it i.e.} the appropriate surface terms associated with the Cotton tensor \eqref{cott}), computed in Appendix \ref{sec:CSsurf}, are of order $\mathcal O(a^2)$ in the (small) angular momentum parameter $a$. As such, there is no back reaction  of the axion field (and hence Cotton tensor) on the wormhole thin shell.}

\section{Conclusions and Outlook \label{sec:concl}}

In this work we studied the scalarization of a slowing rotating black hole in the presence of an axion field coupled to  Chern-Simons term. We found an exact slow rotating Kerr-type black hole solution dressed with axionic hair. Our solution is expressed in terms of the  inverse powers of the radial distance from the centre of the black hole. This behaviour of the black hole solution allows us to go arbitrarily close to the horizon of the hairy black hole. In this way we were able to study the properties of the axionic matter distribution outside the slow rotation Kerr-type black hole.

We  discussed in detail the violation of energy conditions that allowed  the existence of secondary axion hair of our slow rotating Kerr-type black hole solution. There is  strong relation between the energy conditions and the formation of the axionic  hair distribution outside the horizon of our black hole solution. We found that because of the presence of the axion coupling to the Chern-Simons term the null energy condition is violated,  while the violation becomes more important as the coupling becomes stronger and  the violation reaches its maximum near the horizon and then it tends to zero. Thus, we can conclude that the coupling with the curvature term deforms the geometry around the black hole in such a way that the axion hair is allowed.

Motivated by the violation of the null energy condition, we studied the generation of a wormhole, by the gluing   procedure, that connects via a wormhole throat our two  slowly rotating Kerr-type  black holes  with secondary axion hair. We found that the condition for the axion field to not contain any discontinuities on the throat, enforces the two black hole solutions to be counter-rotating. For the slowly rotating to first order in the angular parameter $\mathcal O(a)$ our  black hole solution, we did not find any backreaction of the Cotton tensor on the thin-shell of the wormhole.

It would be interesting to extent our study to higher order $\mathcal O(a)$ in the  angular momentum. We expect to find a much richer structure of the axionic hairy Kerr-type black hole. Then calculating the energy conditions we will get a better understanding of the interplay of the axionic matter  to curvature outside the black hole horizon. Also we expect to find that there will be a backreaction of the Chern-Simons term on the throat of the wormhole and then it will be possible to test  the stability of the resulting thin-shell wormhole.

\section*{Acknowledgements}
We wish to thank Ioannis Kavvadias for fruitful discussions related to material presented in  Appendix \ref{sec:gamma}. The
work of N.E.M is supported in part by the UK Science
and Technology Facilities research Council (STFC) under
the research grant ST/T000759/1. N.E.M.  also acknowledges participation
in the COST Association
Action CA18108 \textit{Quantum Gravity Phenomenology in the Multimessenger Approach (QG-MM)}.

\appendix \section*{APPENDICES}

\section{Mathematical properties of the Chern-Simons action and the Cotton tensor }
\label{sec:Cotton}

In our analysis here we follow the Lorentzian signature, where the metric has one negative eigenvalue, $(-,+,+,+)$ and the Riemann tensor is expressed as
\begin{align}\label{Riemann}
R^{\rho}_{\,\,\,\sigma\mu\nu}=\partial_\mu \Gamma^{\rho}_{\,\,\,\sigma\nu}+\Gamma^{\rho}_{\,\,\,\xi\mu}\Gamma^{\xi}_{\,\,\,\sigma\nu} - \dots \, .
\end{align}

The Cotton tensor, in the way defined in \cite{jackiw}, follows from the
infinitesimal variation with respect to the graviton field of the gravitational anomaly CS term in the action \eqref{eq:Action} with respect to
 $ g_{\mu\nu}$. Using the definitions and conventions of the Riemann tensor \eqref{Riemann} and its dual,
$$\widetilde{R}^{\alpha}_{\,\,\,\beta\gamma\delta}=\frac{1}{2}R^{\alpha}_{\,\,\,\beta\rho\sigma}\varepsilon^{\rho\sigma}_{\,\,\,\,\,\,\gamma\delta}\,,$$
where $\varepsilon_{\rho\sigma\kappa\lambda}$ is the covariant Levi-Civita tensor under the convention $\varepsilon_{0123}=\sqrt{-g}$, which implies $\displaystyle \varepsilon^{0123}=\frac{-1}{\sqrt{-g}}$,
we obtain
\begin{align} \label{Cottder}
\delta  \, \Big( A\,  \int d^4 x  \, \sqrt{-g} \, b(x) \, R^{\mu\nu\rho\sigma} (x) \, \widetilde R_{\mu\nu\rho\sigma}(x) \Big)
\equiv 4\, A \, \int d^4 x \,  \sqrt{-g}\, C^{\mu\nu}(x)\, \delta g_{\mu\nu}(x)
= - 4\, A \, \int d^4 x \,  \sqrt{-g}\, C_{\mu\nu}(x)\, \delta g^{\mu\nu}(x)~,
\end{align}
with $\delta g_{\mu\nu}$ denoting the variation of the covariant metric tensor, and we defined the variation $\delta$ of the CS terms with respect to it. The quantity
$C_{\mu\nu}$ is the Cotton tensor given by
\begin{equation}\label{cottdef2}
	C_{\mu\nu}=-\frac{1}{2}\nabla^{\alpha}\left[(\nabla^{\beta} b) \widetilde{R}_{\alpha\mu\beta\nu}+(\nabla^{\beta} b) \widetilde{R}_{\alpha\nu\beta\mu}\right]~.
\end{equation}
In arriving at \eqref{Cottder} we took into account the infinitesimal gravitational variation of the Riemann tensor \eqref{Riemann},
the Bianchi identity $\varepsilon^{\beta\alpha\rho\sigma} \, \nabla_\alpha \, R_{\rho\sigma}^{\,\,\,\,\,\,\,\mu\nu} =0$,
and ignored surface terms at infinity, arising from appropriate integrations by parts in \eqref{Cottder}.

From the form of the Cotton tensor  \eqref{cottdef2}, it is straightforward to derive that it is traceless, $g^{\mu\nu}\, C_{\mu\nu} = 0~$
\begin{align}\label{trace}
	&C_{\mu\nu}=-\frac{1}{4}\nabla_{\alpha}\left[(\nabla^{\beta}b)(R^{a}_{\,\,\,\mu\gamma\delta}\varepsilon^{\gamma\delta}_{\,\,\,\,\,\,\beta\nu}+R^{a}_{\,\,\,\nu\gamma\delta}\varepsilon^{\gamma\delta}_{\,\,\,\,\,\,\beta\mu})\right]\nonumber \\
	&\implies g^{\mu\nu}C_{\mu\nu}=\frac{1}{2}\nabla_{\alpha}\left[(\nabla_{\beta}b)(R^{a}_{\,\,\,\mu\gamma\delta})\varepsilon^{\mu\gamma\delta\beta}\right]=0~.
\end{align}
The last equality stems from the Riemann tensor symmetries. In particular, recall that $R^{\alpha}_{\,\,\,[\mu\nu\rho]}=0$,
where $[\dots]$ denotes complete antisymmetrization of the indices,
which immediately implies
\begin{equation} R^{\alpha}_{\,\,\,\kappa\lambda\xi}\, \varepsilon^{\kappa\lambda\xi\sigma}=0~.
\end{equation}

A second important property is that
 its covariant derivative satisfies \eqref{Cottcons}, which we prove below, for completeness
\begin{align*}
\nabla_\mu C^{\mu\nu}&=-\frac{1}{2}\nabla_\mu\nabla_\rho\left[(\nabla_\sigma b)(\widetilde{R}^{\rho\mu\sigma\nu}+\widetilde{R}^{\rho\nu\sigma\mu})\right]\\
	&=-\frac{1}{2}\nabla_\mu\nabla_\rho\left[(\nabla_\sigma b)\widetilde{R}^{\rho\mu\sigma\nu}\right]-\frac{1}{2}\nabla_\mu\nabla_\rho\left[(\nabla_\sigma b)\widetilde{R}^{\rho\nu\sigma\mu}\right]\\
	&=-\frac{1}{2}\left(\nabla_{[\mu}\nabla_{\rho]}\right)\left[(\nabla_\sigma b)\widetilde{R}^{\rho\mu\sigma\nu}\right]-\frac{1}{2}(\nabla_\mu\nabla_\rho-\nabla_\rho\nabla_\mu)\left[(\nabla_\sigma b)\widetilde{R}^{\rho\nu\sigma\mu}\right]-\frac{1}{2}\nabla_\rho\nabla_\mu\left[(\nabla_\sigma b)\widetilde{R}^{\rho\nu\sigma\mu}\right]\\
	&=-\frac{1}{2}(\nabla_\mu\nabla_\rho-\nabla_\rho\nabla_\mu)\left[(\nabla_\sigma b)(\widetilde{R}^{\rho\nu\sigma\mu}+\frac{1}{2}\widetilde{R}^{\rho\mu\sigma\nu})\right]\\
	&=-\frac{\nabla_\sigma b}{2}\left[-R_{\xi\mu}\left(\widetilde{R}^{\xi\nu\sigma\mu}+\frac{\widetilde{R}^{\xi\mu\sigma\nu}}{2}\right)+R^{\nu}_{\,\,\,\xi\mu\rho}\left(\widetilde{R}^{\rho\xi\sigma\mu}+\frac{\widetilde{R}^{\rho\mu\sigma\xi}}{2}\right)+R_{\xi\rho}\left(\widetilde{R}^{\rho\nu\sigma\xi}+\frac{\widetilde{R}^{\rho\xi\sigma\nu}}{2}\right)\right]\\
	&=-\frac{\partial_\sigma b}{2}\left[R^{\nu}_{\,\,\,\xi\mu\rho}\left(\widetilde{R}^{\rho\xi\sigma\mu}+\frac{\widetilde{R}^{\rho\mu\sigma\xi}}{2}\right)\right]
	=-\frac{\partial_\sigma b}{4}\left[\widetilde{R}^{\rho\xi\sigma\mu}(R^{\nu}_{\,\,\,\xi\mu\rho}-R^{\nu}_{\,\,\,\rho\mu\xi})+R^{\nu}_{\,\,\,\xi\mu\rho}\widetilde{R}^{\rho\mu\sigma\xi}\right]\\
	&=-\frac{\partial_\sigma b}{4}\left[\widetilde{R}^{\rho\xi\sigma\mu}R^{\nu}_{\,\,\,\mu\xi\rho}+R^{\nu}_{\,\,\,\xi\mu\rho}\widetilde{R}^{\rho\mu\sigma\xi}\right]
	=-\frac{\partial_\sigma b}{2}\widetilde{R}^{\rho\xi\sigma\mu}R^{\nu}_{\,\,\,\mu\xi\rho}
	=-\frac{\partial_\sigma b}{2}\widetilde{R}^{\rho\,\,\,\sigma\mu}_{\,\,\,\xi}R^{\xi\,\,\,\nu}_{\,\,\,\rho\,\,\,\mu}
	=-\frac{1}{4}(\partial^{\nu}b) R_{CS}~.
\end{align*}
where $R_{CS}$ is defined in \eqref{RCS}, and in the last line we used the identity
\begin{equation}
	\widetilde{R}^{\rho\,\,\,\sigma\mu}_{\,\,\,\xi}R^{\xi\,\,\,\nu}_{\,\,\,\rho\,\,\,\mu}=\frac{1}{4}g^{\sigma\nu}\widetilde{R}^{\rho\,\,\,\lambda\mu}_{\,\,\,\xi}R^{\xi}_{\,\,\,\rho\lambda\mu}= \frac{1}{2}g^{\sigma\nu} R_{CS}~.
\end{equation}

\section{Solution of the differential equation (\Ref{eq:firstDE}) \label{sec:appDEsol}}

In this Appendix, we consider the solution of the differential equation (\Ref{eq:firstDE}). Naturally, the first step is to solve the homogeneous term of the ODE,
\begin{equation}
	-2 u(r)+2(r-M)u^\prime(r)+(r^2-2 M r)u^{\prime\prime}(r) =0~.
	\label{A1}
\end{equation}
We note that a particular solution is $\displaystyle u_1=c_1\left(\frac{r-M}{M}\right)$. We may use this solution to simplify the ODE via $u=u_1(r) z(r)$. Then, our simplified expression reads
\begin{equation}
	\label{A2}
	\frac{1}{M}\left[2(M^2-4Mr+2r^2)z'+r(2M^2-3Mr+r^2)z''\right]=0~,
\end{equation}
which implies that
\begin{align}
	\nonumber	
		(\ln z')&=\int \frac{8Mr-2M^2-4r^2}{(r-2M)(r-M)r}dr=-\ln[r(r-2M)(r-M)^2]+c_2\\
		\nonumber
	\implies	z'&=\frac{c_2}{r(r-2M)(r-M)^2}\\
	\nonumber
	z&=\frac{c_2}{(r-M)M^2}+\frac{c_2}{2M^3}\ln\left[1-\frac{2M}{r}\right]+c_3~,
\end{align}
which means that the homogeneous solution to the ODE is
\begin{equation}
	\label{A3}
	u_h=c_1 u_1(r)+c_2 u_2(r)~,
\end{equation}
where
\begin{equation}
	\label{A4}
	u_1(r)=\left(\frac{r-M}{M}\right)~,\qquad u_2(r)=1+\frac{1}{2M}(r-M)\ln\left[1-\frac{2M}{r}\right]~.
\end{equation}
In order to find the complete solution, we will make use of the method of variation of parameters. We consider that the general solution of the differential equation is expressed as
\begin{equation}
	\label{A5}
	u(r)=C_1(r)u_1(r)+C_2(r)u_2(r)~.
\end{equation}
Then, $C_1(r),C_2(r)$ can be solved by the system
\begin{align}
	\label{A6}
&	C'_1(r)u_1(r)+C'_2(r)u_2(r)=0~,\\
	\label{A7}
&	C'_1(r)u'_1(r)+C'_2(r)u'_2(r)=\frac{144 M^2}{r^5(r^2-2M r)}~.
\end{align}
Therefore,
\begin{equation}
	\label{A8}
	\begin{bmatrix}
		C_1' \\ C_2'
	\end{bmatrix}
	=\frac{1}{\mathcal{W}}
	\begin{bmatrix}
		u_2' & -u_2\\
		-u_1'& u_1
	\end{bmatrix}
	\begin{bmatrix}
		0 \\ \frac{144 M^2}{r^5(r^2-2M r)}
	\end{bmatrix}~,
\end{equation}
where $\mathcal{W}$ denotes the Wronskian of our solutions. The system can be easily solved to yield that

\begin{align}
	\label{A9}
	C_1(r)&=\frac{81 M}{2r^4}-\frac{5}{r^3}-\frac{15}{4Mr^2}-\frac{15}{4 M^2 r}+\ln\left(1-\frac{2M}{r}\right)\frac{6 (4r-3M)}{r^4}-\frac{15}{8M^3}\ln\left(1-\frac{2M}{r}\right)+c_1~,\\
	\label{A10}
	C_2(r)&=\frac{36M}{r^4}-\frac{48}{r^3}+c_2~.
\end{align}
Making use of (\Ref{A5}), we find that the complete solution reads
\begin{equation}
	\label{A11}
	u(r)=-c_1+c_2-\frac{15}{4M^3}-\frac{9M}{2r^4}-\frac{5}{2r^3}-\frac{5}{4Mr^2}+c_1\frac{r}{M}+\ln\left(1-\frac{2M}{r}\right)\left[\frac{15}{8M^3}-\frac{c_2}{2}\right]+\ln\left(1-\frac{2M}{r}\right)\left[\frac{c_2 r}{2M}-\frac{15 r}{8M^4}\right]~.
\end{equation}
In order to cancel the divergent terms, we fix the integration constants to $c_1=0$ and $\displaystyle c_2=\frac{15}{4M^3}$ and we find that the asymptotic solution of the axion reads
\begin{equation}
	\label{A12}
	u(r)=-\frac{9M}{2r^4}-\frac{5}{2r^3}-\frac{5}{4M r^2}~.
\end{equation}

\section{Proof of Convergence of the series \eqref{sumcor}}\label{sec:gamma}

The relevant series appears in the $g_{t\phi}$ element of the metric of our black hole solution \eqref{hth},  see \eqref{offdiag}, \eqref{sumcor}:
\begin{equation}
	g_{t\phi}= r^2\, \left(-\frac{2 M}{r^3}-w(r)\right)a\;\sin^2(\th) \equiv \left(-\frac{2 M}{r}-\widetilde w(r)\right)a\;\sin^2(\th), \quad 
	\widetilde w(r)=\sum_{n=4}^{\infty}\frac{d_n M^{n-2}}{r^{n-2}}\,,
	\label{tphicomponentsolution}
\end{equation}
where $d_n$ is determined by the following recurrence relation \eqref{dnequation}:
\begin{equation}
	d_n=\frac{2(n-5)^2(n-1)}{n(n-6)(n-3)}d_{n-1}+\frac{576\gamma^2}{n(n-3)}d_{n-6},\;\;\text{for}\;n\geq 10~,
	\label{dnequation1}
\end{equation}
and initial conditions \eqref{ddef}: 
\begin{equation}\label{ddef2}
	d_4=d_5=0\;\;,\;\;d_6=-5\g^2\;\;,\;\;d_7=-\frac{60\g^2}{7}\;\;,\;\;d_8=-\frac{27\g^2}{2}\;\;,\;\;d_9=0
\end{equation}
with $\gamma \in \mathbb R$. We restrict ourselves in the exterior to horizon region  $r\geq 2M$.

For notational convenience and brevity we redefine $r \rightarrow r/M$, hence
\begin{equation}\label{sumd}
	\widetilde w(r)=\sum_{n=4}^{\infty}\frac{d_n}{r^{n-2}}.
\end{equation}

Below we shall prove the convergence of this sum for all $r\geq 2$ for all $\gamma \in \mathbb R$. 

To this end, we first note that $d_n\leq0\;, \, \forall n$. Thus, we define the sequence $c_n=-d_n$, for which $c_n\geq0,\;\forall n$. Then, in terms of $c_n$, we have:
\begin{equation}\label{wr}
	\widetilde w(r)=-\sum_{n=4}^{\infty}\frac{c_n}{r^{n-2}}
\end{equation}   
where
\begin{equation}
	c_n=a_n c_{n-1}+b_n c_{n-6},\;\;\text{for}\;n\geq 10~,
	\label{cnequation}
\end{equation}
with
\begin{equation}
		a_n=\frac{2(n-5)^2(n-1)}{n(n-6)(n-3)}\, \quad 
		b_n=\frac{576\gamma^2}{n(n-3)},
\end{equation}
and initial conditions:
\begin{equation}\label{ddef3}
	c_4=c_5=0\;\;,\;\;c_6=5\g^2\;\;,\;\;c_7=\frac{60\g^2}{7}\;\;,\;\;c_8=\frac{27\g^2}{2}\;\;,\;\;c_9=0
\end{equation}

Let us define the sequence: \[\tilde{\Sigma}_N=\sum_{n=4}^{N}\frac{c_n}{r^{n-2}}\] 
As $c_n/r^{n-2}$ are non-negative, $\tilde{\Sigma}_N$ is an increasing sequence, meaning $\tilde{\Sigma}_{N+1}\geq\tilde{\Sigma}_N,\;\forall N\in\mathbb{N}$. \\

For increasing sequences, the following theorem exists \cite{Burkill}:\\

{\bf Theorem:} {\it An increasing sequence tends either to a finite limit or to $+\infty$.}\\

Hence, a necessary and sufficient condition for the convergence of $\tilde{\Sigma}$ is the demonstration that it is bounded, {\it i.e.} that there exists a finite, positive number $\mathcal{N}$, such that:
\[
\tilde{\Sigma}=\sum_{n=4}^{\infty}\frac{c_n}{r^{n-2}}\leq\mathcal{N}
\]

\textbf{\underline{Statement 1}}: \textit{ If $\sum_{n=4}^{\infty}\frac{c_n}{2^{n-2}}$ converges, then $\sum_{n=4}^{\infty}\frac{c_n}{r^{n-2}}\;\text{converges}\;\;\forall r\geq 2$}.\par 
\begin{proof}
Suppose that $\sum_{n=4}^{\infty}\frac{c_n}{2^{n-2}}\leq\mathcal{K}$, where $\mathcal{K}$ finite. For $r>2\rightarrow 1/r^{n-2}<1/2^{n-2} \rightarrow c_n/r^{n-2}\leq c_n/2^{n-2}\;\forall n\geq4$, where the equality holds in the case of $c_n=0$. Thus, 
\[
\begin{aligned}
\sum_{n=4}^{\infty}\frac{c_n}{r^{n-2}}\leq\sum_{n=4}^{\infty}\frac{c_n}{2^{n-2}}\quad 
\Rightarrow  \quad \sum_{n=4}^{\infty}\frac{c_n}{r^{n-2}}\leq\mathcal{K}
\end{aligned}
\]
which means that $\sum_{n=4}^{\infty}\frac{c_n}{r^{n-2}}$ converges $\forall r>2$. \end{proof}
\par 
Hence, according to the above statement 1, we should establish the convergence of the infinite series 
\[
\Sigma=\sum_{n=4}^{\infty}\frac{c_n}{2^{n-2}}
\]

For $a_n$ and $b_n$, we have:
\[
\begin{aligned}
&1)\;\;a_n\rightarrow 2,\;\text{as}\;n\rightarrow +\infty\\
&2)\;\;b_n\rightarrow 0,\;\text{as}\;n\rightarrow +\infty\\
\end{aligned}
\]
i.e. $a_n$ and $b_n$ both converge. 

Below, we use the mathematical result \textit{that a sequence, say $s_n$, is convergent, i.e. tends to a finite limit $s$ as $n\rightarrow \infty$, implies that $s_n$ is bounded} \cite{Burkill}.\par 
This implies that, since $a_n,b_n$ are convergent, in view of the above result, they are bounded, meaning that there exist $k_1,k_2$, such that 
\begin{equation}
\abs{a_n}\leq k_1\;\;and\;\;\abs{b_n}\leq k_2\;\;\forall n\in\mathbb{N}
\end{equation}\par 
\textbf{\underline{Statement 2}}: \textit{The sequence $c_n$ is bounded by induction.}
\begin{proof}
Suppose that there exists a subsequence $c_{N-6},...c_{N-1}$, for some $N$, that is bounded, i.e. there exists $N^\prime$, such that \[\abs{c_{N-6}},\abs{c_{N-5}},...,\abs{c_{N-1}}\leq N^\prime\]. \\
Then, using the triangle inequality
\[
\begin{aligned}
	& \abs{c_N}\leq \abs{a_N}\abs{c_{N-1}}+\abs{b_N}\abs{c_{N-6}} \quad 
	\Rightarrow&\abs{c_N}\leq(k_1+k_2)N^\prime \quad 
	\Rightarrow&\abs{c_N}\leq \widetilde{\mathcal{M}}\,,
\end{aligned}
\]
where $\widetilde{\mathcal{M}}=(k_1+k_2)\, N^\prime$ finite. \\

Thus, as the corresponding subsequence exists for $N=10$, as $c_4,...,c_9$ are finite, by induction, there exists finite $\mathcal{D}$, such that\[\abs{c_n}\leq\mathcal{D},\;\forall n\geq 10\]
concluding that $c_n$ is bounded $\forall n\geq 4$.  
\end{proof}
Thus, as $c_n$ is bounded and non-negative, there exists $\mathcal{D}\,>\,0$, such that:
\[
0\leq c_n\leq \mathcal{D}\;,\forall\;n\geq 4
\] 
Thence, 
\[
\Sigma=\sum_{n=4}^{\infty}\frac{c_n}{2^{n-2}}\leq\mathcal{D}\sum_{n=4}^{\infty}\left(\frac{1}{2}\right)^{n-2}
\]
or
\[
\Sigma\leq\mathcal{D}\sum_{n=2}^{\infty}\left(\frac{1}{2}\right)^n
\]
For the geometric series, we know that
\[
\sum_{n=0}^\infty x^n=\frac{1}{1-x}, \qquad \text{for}\;-1<x<1
\]
Thus, for $x=1/2$, it is easy to show that:
\[
\sum_{n=2}^{\infty}\left(\frac{1}{2}\right)^{n}=\frac{1}{2}
\]
which implies
\[
\Sigma=\sum_{n=4}^{\infty}\frac{c_n}{2^{n-2}}\leq\frac{\mathcal{D}}{2},
\]
i.e., the infinite series $\Sigma$ is bounded. \par

This proves the required result,  
that the sum $\widetilde w(r)$, and thus $w(r)$ (\eqref{sumcor}), converges $\forall r\geq 2$  and $\forall \gamma \in \mathbb R$.

\section{The energy conditions for Einstein spaces }
\label{sec:energcond}

Here we review  the various energy conditions that characterise
gravitational theories in Einstein spacetimes. Suppose that $T_{\m\n}$ refers to the conserved energy-momentum, obeying the Einstein gravitational equations of motion
\begin{equation}
R_{\m\n}=\k^2\left(T_{\m\n}-\frac{1}{2}g_{\m\n}T\right)~.
\label{graveom}
\end{equation}
Notice that if one considers higher-curvature modified gravities, such as scalar-Gauss-Bonnet or axion-Chern-Simons gravities,
the contributions of the modifications are absorbed in the definition of $T_{\m\n}$, viewed as an effective stress tensor.
With this understanding, we review the following energy conditions.

\begin{itemize}

\item Weak Energy Condition (WEC) : The energy density as measured by any observer with a timelike four-velocity $t^\mu$, is non-negative, which formally can be expressed as
\begin{equation}
	T_{\m\n}t^\m t^\n\geq0\;,\;\;\forall\;t:\;\;t^\m t_\m<0~.
	\label{WEC}
\end{equation}

\item Null Energy Condition (NEC): expresses the requirement that the geometry has a focusing (attractive) effect on null geodesics,
\begin{equation}
	T_{\m\n}l^\m l^\n\geq0\;,\;\;\forall\;l:\;\;l^\m l_\m=0~,
	\label{NEC}
\end{equation}
where $l^\m$ is any null four-vector.

We can understand the condition by noting that the Raychadhuri equation describing the focusing of null geodesics is
\[\frac{d}{d\tau}\th_l=-R_{\m\n}l^\m l^\n+...\]
where the parameter $\theta$ in Raychadhuri equation has the interpretation of the rate of change of the congruences' cross sectional volume.
Thus, for a focusing effect on the null geodesics: $R_{\m\n}l^\m l^\n\geq 0$. Hence, the gravitational equations of motion yield the same condition for the conserved energy-momentum tensor.

\item Strong Energy Condition (SEC): expresses the requirement that the geometry has a
focusing (attractive) effect on timelike geodesics.
\begin{equation}
	T_{\m\n}t^\m t^\n\geq \frac{1}{2}T \,g_{\m\n}\,t^\m t^\n\;,\;\;\forall\;t:\;\;t^\m t_\m<0~,
	\label{SEC}
\end{equation}
where $T$ is the gravitational trace of the stress tensor $T \equiv g^{\mu\nu}\, T_{\mu\nu}$.
The condition is understood again by noting that the Raychadhuri equation describing the focusing of timelike geodesics is
\[\frac{d}{d\tau}\th_t=-R_{\m\n}t^\m t^\n+...\]
Thus, for a focusing effect on the timelike geodesics: $R_{\m\n}t^\m t^\n\geq 0$. Hence, the gravitational equations of motion (\ref{graveom}) yield (\ref{SEC}) for the conserved energy-momentum tensor.

 \item Dominant Energy Condition (DEC): This energy condition refers to the current density: $P^\a=-T^\a\!_\b t^\b$, i.e. the energy-momentum current as seen by an observer with 4-velocity $t^\m$. It is essentially the statement that the speed of the flow of energy should not exceed that of light, i.e. $P^\a$ should be causal and future directed for all timelike and future directed $t^\a$. Since $t^\a$ is timelike and future directed, the above conditions are mathematically expressed as follows
  \[
  P_\a t^\a\leq 0\;\;\text{and}\;\;P_\a P^\a\leq 0~,
  \]
  which yields the DEC as a set of the two requirements
  \begin{equation}
T_{\m\n}t^\m t^\n\geq 0\;\;\text{and}\;\;T_{\m\n}T^\m\!_\a t^\n t^\a\leq 0\;\;\forall\;t:\;\;t_\m t^\m<0~.
  \end{equation}
We note that the first of the above is just $T_{\a\b} t^\a t^\b\geq 0$, that is, the WEC.

 \end{itemize}
One can show the relations between the various energy conditions, as a ``hierarchy of the validity'' between them
\begin{equation}
		\begin{aligned}
		& (DEC)\implies (WEC) \implies (NEC)\\
		& (SEC) \implies (NEC)\\
		& \text{(NEC) violation}\;\, {\rm implies} \, \, \text{(DEC), (SEC), (WEC) violation}
	\end{aligned}
\end{equation}
where  the arrow notation $\implies$ implies the validity of a condition.

Thus, the NEC \eqref{NEC} is the weakest energy condition, because if this condition is satisfied, then it is implied  that the rest are saitisfied. However, if we look the relations according to their violation, the NEC is the strongest one, because if this condition is violated all of the other conditions are violated, too. This is the case for the Kerr black-hole solution \eqref{slowKerr}  of our SC-axion gravitational theory  \eqref{eq:Action}~\cite{jackiw,Alexander}, implying the existence of axionic hair~\cite{kerr1,kerr2,kerr3,yunes1,yunes2}.

\section{Computation of the extrinsic-curvature junctioning quantities $\mathfrak{R}^{\alpha}_{\,\,\,\beta\gamma\delta}$
in the wormhole }

\label{sec:curvglue}

The fact that the metric is continuous across $\Sigma$ in the coordinates $x^\mu$ implies that its tangential derivatives also must be continuous. This means that if $\partial_\gamma g_{\alpha\beta}$ is to be discontinuous, the discontinuity must be directed along the normal vector $n^\alpha$.
As such, there exists a tensor field $q_{\alpha\beta}$ such that
\begin{equation}
	\label{B.1}
	[\partial_\gamma g_{\alpha\beta}]=n_\gamma q_{\alpha\beta}\implies q_{\alpha\beta}=\epsilon	[\partial_\gamma g_{\alpha\beta}] n^\gamma~.
\end{equation}
Making use of (\Ref{B.1}) under the first junctioning condition which yields $[g^{\alpha\beta}]=0$, we may find that
\begin{equation}
	\label{B.2}
	[\Gamma^{\alpha}_{\,\,\, \beta\gamma}]=\frac{1}{2}(q^\alpha_{\,\,\,\beta}n_\gamma+q^\alpha_{\,\,\,\gamma}n_\beta-q_{\beta\gamma}n^\alpha)~.
\end{equation}
This means that
\begin{equation}
	\label{B.3}
	\mathfrak{R}^\alpha_{\,\,\,\beta\gamma\delta}=\frac{\epsilon}{2}(q^{\alpha}_{\,\,\,\delta}n_\beta n_\gamma-q^{\alpha}_{\,\,\,\gamma}n_\beta n_\delta-q_{\beta\delta}n^\alpha n_\gamma+q_{\beta\gamma}n^\alpha n_\delta)~.
\end{equation}
Now that we have a tensorial quantity for $\mathfrak{R}$, we may immediately note that the Einstein tensor reads
\begin{equation}
	\label{B.4}
	G_{\alpha\beta}=G^+_{\alpha\beta}\Theta(\lambda)+G^-_{\alpha\beta}\Theta(-\lambda)+\delta(\lambda)\mathfrak{G}_{\alpha\beta}~,
	\end{equation}
where
\begin{equation}
	\label{B.5}
	\mathfrak{G}_{\alpha\beta}=\frac{\epsilon}{2}(q_{\mu\alpha}n^\mu n_\beta+q_{\mu\beta}n^\mu n_\alpha-q n_\alpha n_\beta-\epsilon q_{\alpha\beta}-(q_{\mu\nu}n^\mu n^\nu-\epsilon q)g_{\alpha\beta})~.
\end{equation}
From the form of $\mathfrak{G}_{\alpha\beta}$, we note that it is symmetric, as expected, and that it is also tangent to the hypersurface, since $\mathfrak{G}_{\alpha\beta}n^\alpha=0$. Therefore, it admits the decomposition of
\begin{equation}
	\label{B.6}
	\mathfrak{G}^{\alpha\beta}=\mathfrak{G}^{ij}e^\alpha_i e^\beta_j~.
\end{equation}
 Evaluating the decomposition we find
\begin{align*}
 \mathfrak{G}_{ij}&=\mathfrak{G}_{\alpha\beta}e^\alpha_i e^\beta_j\\
 				&=\frac{\epsilon}{2}(-\epsilon q_{\alpha\beta}e^{a}_i e^\beta_j+\epsilon q_{\mu\nu}(g^{\mu\nu}-\epsilon n^\mu n^\nu)h_{ij})\\
 				&=\frac{1}{2}(- q_{\alpha\beta}e^{a}_i e^\beta_j+ q_{\mu\nu}h^{\mu\nu}h_{ij})~.
\end{align*}
We now note that $q_{\alpha\beta}e^\alpha_ie^\beta_j$ is just the jump of the extrinsic curvature across the hypersurface. Indeed, calculating
\begin{align*}
	[K_{ij}]&=[(\nabla_\alpha n_{\beta}) e^\alpha_i e^\beta_j]\\
	&\overset{(\Ref{w.4})}{=}[(\nabla_\alpha n_{\beta})] e^\alpha_i e^\beta_j\\
	&=-\left([\Gamma^{\gamma}_{\,\,\, \alpha\beta}]n_\gamma\right)e^\alpha_i e^\beta_j\\	
	&\overset{(\Ref{B.2})}{=}-\left(\frac{1}{2}(q^\gamma_{\,\,\,\beta}n_\alpha+q^\gamma_{\,\,\,\alpha}n_\beta-q_{\alpha\beta}n^\gamma)\right)n_\gamma e^\alpha_i e^\beta_j\\
	&=\frac{\epsilon}{2}q_{\alpha\beta}e^\alpha_i e^\beta_j~,
\end{align*}
we find that
\begin{equation}
	\label{B.7}
	[K_{ij}]=\frac{\epsilon}{2}q_{\alpha\beta}e^\alpha_i e^\beta_j,\qquad [K]=\frac{\epsilon}{2}h^{mn}(q_{\alpha\beta}e^\alpha_me^\beta_n)~.
\end{equation}
Therefore, since $K_{\mu\nu}n^\mu=0$, i.e. the extrinsic curvature of the hypersurface is tangent to the hypersurface, as is well known, we find by the decomposition $K_{ij}=K_{\mu\nu}e^\mu_i e^\mu_j$ that
\begin{equation}
	\label{B.8}
	q_{\alpha\beta}=\frac{2}{\epsilon}[K_{\alpha\beta}]~,
\end{equation}
which yields that
\begin{equation}
	\label{B.9}
	\mathfrak{R}^{\alpha}_{\,\,\,\beta\gamma\delta}=(n_\beta n_\gamma [K^{\alpha}_{\delta}]-n_\beta n_\delta [K^\alpha_\gamma]-[K_{\beta\delta}]n^\alpha n_\gamma+[K_{\beta\gamma}]n^\alpha n\delta)~,
\end{equation}
which was used in the text.

\section{Chern-Simons (surface) contributions on the Wormhole thin shell }
\label{sec:CSsurf}

We consider the CS term of the action \eqref{eq:Action}
\begin{equation}
	I_{CS}=\int d^4x\, \sqrt{-g}\, b R\widetilde{R}~.
\end{equation}
Upon taking the variation of the $b R\widetilde{R}$ term we find that
\begin{equation}
	\delta I_{CS} \sim2 \int d^4x\, \sqrt{-g}\, b R^{\alpha}_{\,\,\,\beta \kappa\lambda}\varepsilon ^{\kappa\lambda\rho\sigma}(\nabla_\rho \delta \Gamma^{\beta}_{\,\,\,\alpha\sigma})~,
\end{equation}
whose surface term is
\begin{equation}
	\int d\Sigma_\rho b R^{\alpha}_{\,\,\,\beta \kappa\lambda}\varepsilon ^{\kappa\lambda\rho\sigma}(\delta \Gamma^{\beta}_{\,\,\,\alpha\sigma})~.
\end{equation}
Recalling that $\delta \Gamma$ reads under $g^{\mu\nu}\rightarrow g^{\mu\nu}+\delta g^{\mu\nu}$,
\begin{equation}
	\delta \Gamma^\beta_{\,\,\,\alpha\sigma}=-\frac{1}{2}\left[g_{\mu\sigma}\nabla_\alpha(\delta g^{\mu\beta})+g_{\mu\alpha}\nabla_\sigma(\delta g^{\mu\beta})-g_{\alpha\nu}g_{\sigma\mu}\nabla^\beta(\delta g^{\mu\nu})\right]~,
\end{equation}
we have that
\begin{align*}
	\int d\Sigma_\rho b R^{\alpha}_{\,\,\,\beta \kappa\lambda}\varepsilon ^{\kappa\lambda\rho\sigma}(\delta \Gamma^{\beta}_{\,\,\,\alpha\sigma})&=-\int d\Sigma_\rho b \widetilde{R}^{\alpha\,\,\,\rho\sigma}_{\,\,\,\beta}g_{\mu\sigma}\nabla_\alpha (\delta g^{\mu\beta})-\int d\Sigma_\rho b \widetilde{R}^{\alpha\,\,\,\rho\sigma}_{\,\,\,\beta}g_{\mu\alpha}\nabla_\sigma (\delta g^{\mu\beta})+\int d\Sigma_\rho b \widetilde{R}^{\alpha\,\,\,\rho\sigma}_{\,\,\,\beta}g_{\alpha\mu}g_{\sigma\nu}\nabla^\beta (\delta g^{\mu\nu})\\
	&=-\int d\Sigma_\rho b \widetilde{R}^{\alpha\,\,\,\rho}_{\,\,\,\beta\,\,\,\mu}\nabla_\alpha (\delta g^{\mu\beta})-\int d\Sigma_\rho b \widetilde{R}^{\,\,\,\,\,\,\rho\alpha}_{\mu\beta}\nabla_\alpha (\delta g^{\mu\beta})+\int d\Sigma_\rho b \widetilde{R}^{\,\,\,\alpha\rho}_{\mu\,\,\,\,\,\,\beta}\nabla_\alpha (\delta g^{\mu\beta})\\
	&=\int d\Sigma_\rho b (\widetilde{R}^{\,\,\,\alpha\rho}_{\mu\,\,\,\,\,\,\nu}-\widetilde{R}^{\alpha\,\,\,\rho}_{\,\,\,\nu\,\,\,\mu})\nabla_\alpha (\delta g^{\mu\nu})\\
	&=\int d^3x\sqrt{h}n^\rho b (\widetilde{R}_{\mu\beta\rho\nu}+\widetilde{R}_{\nu\beta\rho\mu})g^{\alpha\beta}\nabla_\alpha (\delta g^{\mu\nu})~.\\
\end{align*}
Since $\delta g^{\mu\nu}$ is zero on $\Sigma$, $\nabla_\alpha \delta g^{\mu\nu}=\partial_\alpha \delta g^{\mu\nu}$. Also, by the decomposition of $g^{\alpha\beta}=h^{\alpha\beta}+\epsilon n^\alpha n^\beta$, the tangential derivative of the variation of the metric also vanishes on $\Sigma$. Therefore,
\begin{align*}
	\int d\Sigma_\rho b R^{\alpha}_{\,\,\,\beta \kappa\lambda}\epsilon ^{\kappa\lambda\rho\sigma}(\delta \Gamma^{\beta}_{\,\,\,\alpha\sigma})&=\int d^3x\sqrt{h}n^\rho b (\widetilde{R}_{\mu\beta\rho\nu}+\widetilde{R}_{\nu\beta\rho\mu})\epsilon  n^\beta n^\alpha \partial_\alpha (\delta g^{\mu\nu})~.\\
\end{align*}
Computing the interior of the integral for our case, it is found that it is of order $\mathcal{O}(a)^2$ or higher, without even using the constraint to calculate it on the hypersurface.

\end{document}